\pdfoutput=1
\documentclass[usenatbib,useAMS]{mnras}
\usepackage{txfonts}
\usepackage{graphicx}
\usepackage{dcolumn}

\begin{document}

\title[Dust grains in turbulent molecular clouds]{Clustering and dynamic decoupling of dust grains in turbulent molecular clouds}

\author[Mattsson et al.]{Lars Mattsson$^{1}$\thanks{E-mail: lars.mattsson@nordita.org},  
Akshay Bhatnagar$^1$,
Fred~A. Gent$^2$,
Beatriz Villarroel$^{1,3}$\\
$^1$Nordita, KTH Royal Institute of Technology and Stockholm University, Roslagstullsbacken 23, SE-106 91 Stockholm, Sweden\\
$^2$ReSoLVE Centre of Excellence, Department of Computer Science, Aalto University, PO Box 15400, FI-00076 Aalto, Finland\\
$^3$Department of Information Technology, Uppsala University, Box 337, SE-751 05 Uppsala, Sweden}

\pagerange{\pageref{firstpage}--\pageref{lastpage}} 
\pubyear{2018}

\maketitle

\label{firstpage}

\date{\today}

\begin{abstract} 
We present high resolution ($1024^3$) simulations of super-/hyper-sonic isothermal hydrodynamic turbulence inside an interstellar molecular cloud (resolving scales of typically 20 -- 100 AU), including a multi-disperse population of dust grains, i.e., a range of grain sizes is considered. 
Due to inertia, large grains (typical radius $a \gtrsim 1.0\,\mu$m) will decouple from the gas flow, while small grains ($a\lesssim 0.1\,\mu$m) will tend to better trace the motions of the gas. We note that simulations with purely solenoidal forcing show somewhat more pronounced decoupling and less clustering compared to simulations with purely compressive forcing. Overall, small and large grains tend to cluster, while intermediate-size grains show essentially a random isotropic distribution. As a consequence of increased clustering, the grain-grain interaction rate is locally elevated; but since small and large grains are often not spatially correlated, it is unclear what effect this clustering would have on the coagulation rate. Due to spatial separation of dust and gas, a diffuse upper limit to the grain sizes obtained by condensational growth is also expected, since large (decoupled) grains are not necessarily located where the growth species in the molecular gas is. 
\end{abstract}


\section{Introduction}
Dust is an important constituent of the interstellar medium (ISM) of the Galaxy and the main source of opacity and light scattering. The rarefied gas of the ISM is also highly turbulent, which make cosmic dust a perfect example of inertial particles in a turbulent flow; a classical problem in fluid mechanics. A crucial difference is that in classical fluid dynamics one considers incompressible flows, while in the ISM compressibility really matters. Regardless of how interstellar turbulence is induced and maintained, the hypersonic\footnote{Hypersonic flow is here defined as a flow with a characteristic Mach number, i.e., the ratio of the root-mean-square gas velocity and the sound speed $\mathcal{M}_{\rm rms}=u_{\rm rms}/c_{\rm s}$, which is between 5 and 25.} nature of interstellar turbulence,  evidenced by large non-thermal line widths  \citep{Larson81,Solomon87}, suggests that the Stokes number, defined as the ratio  the time it takes for a stationary particle to couple to the flow over the characteristic flow timescale, is relatively high for cosmic dust grains. This is the case despite the small sizes of the grains, which is due to the fact that the Stokes number is proportional to the Mach number (which is high).  Grains with radii $\sim 1\,\mu$m are thus expected to decouple from the gas and there is plenty of evidence for the existence of such grains in molecular clouds (MCs) in the ISM \citep[e.g.,][]{Pagani10,Steinacker10,Steinacker15,Ysard16,Saajasto18}. \citet{Hopkins16} presented a suite of simulations of the turbulent dynamics of an MC including a dust phase. Due to the dynamical decoupling between gas and dust grains (i.e., a relatively long kinetic-drag timescale) the gas and dust do not necessarily end up at the same place, which results in a wide distribution of dust-to-gas ratios, according to \citet{Hopkins16}.  Similar results have been found in several other studies \citep[e.g.,][]{Padoan06,Downes12} but see also \citet{Tricco17}. It is well established that turbulence can cause increased concentrations of dust particles, leading to an increased rate of particle-particle interactions and thus enable coagulation on sufficiently short timescales \citep{Pumir16}. But the spatial separation of dust and gas should also have an effect on the rate of dust formation by condensation as the number of gas molecules hitting the surfaces of the dust grains per unit time would be lower.   

Interstellar grain growth by condensation is considered to be an important dust-formation channel, not the least as a necessary replenishment mechanism to counteract dust destruction in the ISM. Indirect support to the interstellar-growth hypothesis comes from destruction of grains in the ISM, highlighting a need for a replenishment mechanism \citep[see, e.g.,][]{McKee89,Draine90,Ginolfi18} -- a need which appears to be even greater in the early universe \citep{Mattsson11b,Valiante11}. The observed depletion patterns in interstellar gas are also consistent with dust depletion due condensation in MCs \citep{Savage96a,Savage96b,Jenkins09,DeCia16}. Further indirect evidence comes from the fact that late-type galaxies seem to have steeper dust-to-gas gradients than metallicity gradients along the radial extension of their discs \citep{Mattsson12a,Mattsson12b,Mattsson14b}. If gas-dust separation due to turbulence is as important as it seems to be, there could very well be a significant damping of the growth rate due to turbulent gas motions, once the grains have a certain size. At the same time, the high-density peaks created by turbulence must increase the condensation rate, at least locally. Dust growth in MCs may therefore be both limited and boosted by turbulence. The rate of coagulation (and shattering) is also boosted by turbulence due to locally increased number densities and relative particle speeds. Understanding how turbulence shapes the size distribution of dust grains, and consequently affects the grain temperature distribution, is important for our general understanding of cosmic dust, not least because it may have significant effects on the infrared flux-to-mass ratio \citep[see][and references therein]{Mattsson15}.

Interstellar high Mach-number turbulence is undoubtedly forced turbulence in one way or another. Thus, simulation of the ISM requires using a forcing scheme. The driving force behind interstellar turbulence is not fully understood, but it is widely assumed that the kinetic-energy injection by supernovae (SNe) must play an important role \citep{Elmegreen04,Schmidt09,Padoan16,Pan16,Padoan16b}. The velocity field usually shows only a weak vorticity component \citep{Elmegreen04}, which seems natural in case interstellar turbulence is driven by SN shocks. Moreover, the interstellar gas flow is highly compressible and must have a very high Reynolds number. But in the coldest and densest phase of the ISM, the MCs, there may be a significant vorticity component  and it must be considered unclear whether the energy injection from SN shockwaves can drive the turbulence on spatial scales of MCs. Very recent work by \citet{Seifried18} seem to show that the SN-driven turbulence is in fact a problematic hypothesis. Although the energy released from SNe has been demonstrated to be sufficient to drive interstellar turbulence \citep{Vazquez94}, and also more recent simulations seem to support SN driven turbulence in the diffuse ISM \citep[see, e.g.,][]{Korpi99,Mee06,Brandenburg07,Padoan16, Gent13a,Gent13b}, the impact of the SN shocks inside MCs may not be that large. 

In the present paper we investigate the clustering and dynamic decoupling of dust grains expected in a MC (cold ISM) using direct numerical simulations of stochastically forced, supersonic/hypersonic isothermal hydrodynamic turbulence. Compared to many previous studies, our investigation adds the following: 
\begin{itemize}
\item[(1)] three-dimensional simulations in high resolution ($1024^3$); 
\item[(2)] a multi-disperse dust component, i.e., a range of grain sizes is followed simultaneously\footnote{We have chosen to focus to somewhat larger grain than \citet{Hopkins16}, because we want to adress the conflicting results of their study and that of \citet{Tricco17}, where the latter is suggesting that only very large grains decouple from the gas flow.}; 
\item[(3)] comparison of particle dynamics in turbulent flows with compressive and solenoidal (rotational) forcing; 
\item[(4)] quantitative analysis of the clustering of grains due to turbulent dynamics and velocity decoupling between the gas and dust phases.
\end{itemize}

This paper is organised as follows. Section 2 gives a general background on the method of simulation and underlying physical theory. In section 3 we present results and analysis. We also discuss the implications for clustering  and subsequent processes of grains in MCs. In section 4 we summarise our findings and future outlook.

\begin{figure}
    \resizebox{\hsize}{!}{ 
    \includegraphics{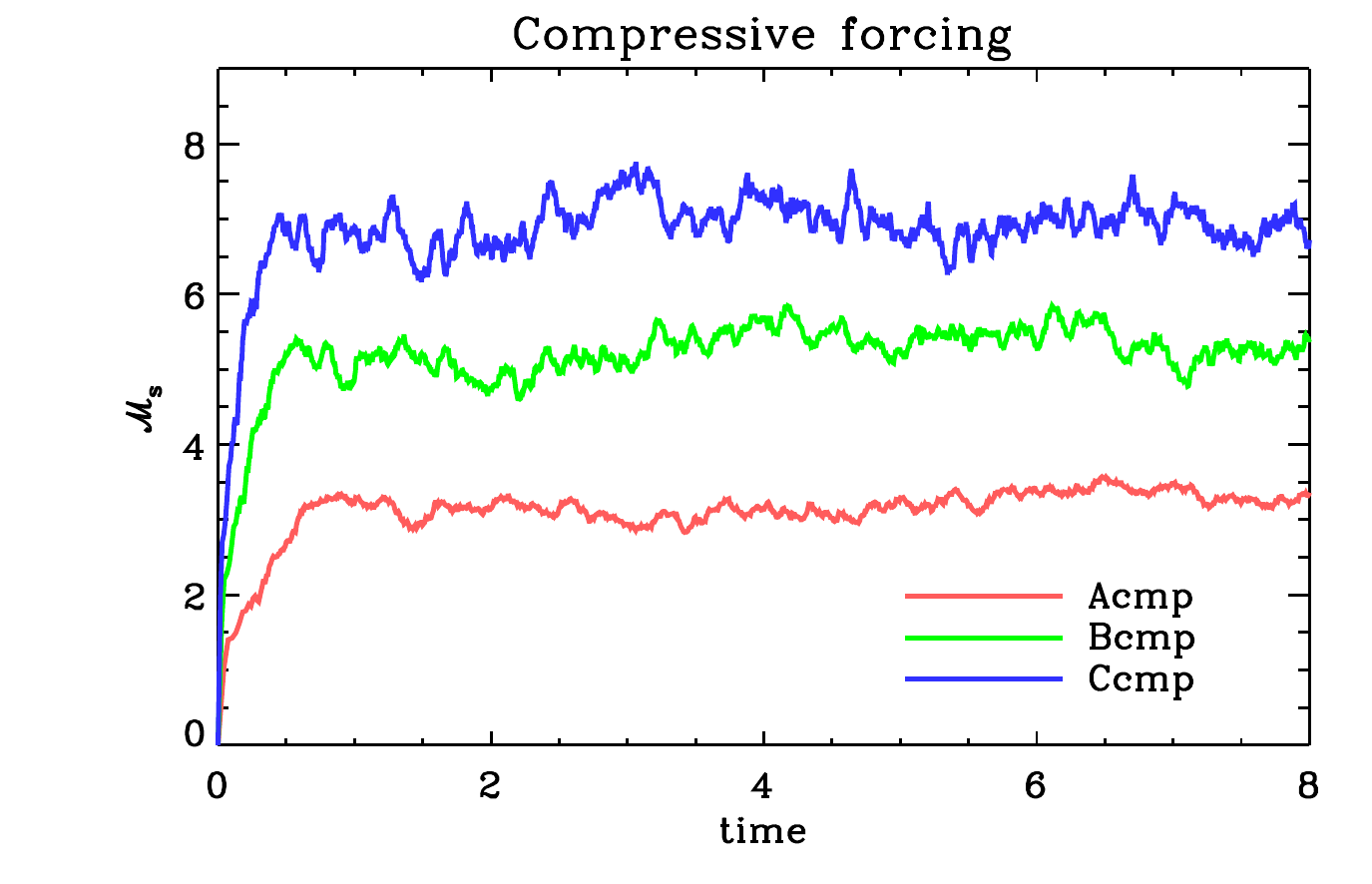}}
    \resizebox{\hsize}{!}{ 
    \includegraphics{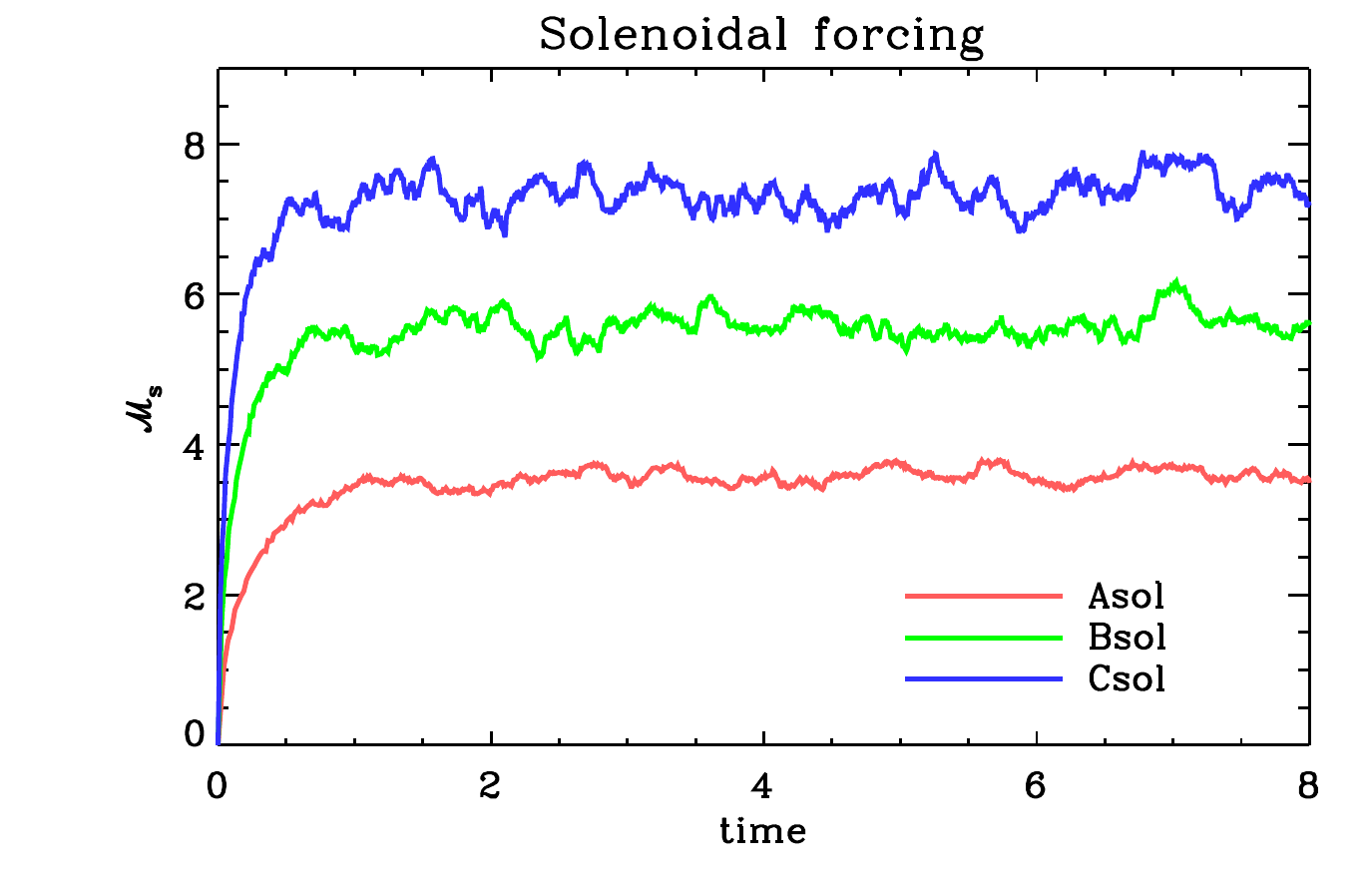}}

  \caption{\label{timeseries} Upper panel: time evolution for the root-mean-square Mach number ($\mathcal{M}_{\rm rms} = u_{\rm rms}/c_{\rm s}$) for simulations with purely compressive forcing. Lower panel: the same as the left panel, but for simulations with purely solenoidal forcing.}
\end{figure}

        \begin{table*}
  \begin{center}
  \caption{\label{simulations} Properties and physical parameters of the simulations. All simulations have the mean gas density and isothermal sound speed set to unity, i.e., $\langle\rho\rangle=c_{\rm s}=1$.}
  \begin{tabular}{l|llllllll}
  \hline
  \hline
  \rule[-0.2cm]{0mm}{0.8cm}
  Simulation & $f$  &$\langle\log(\rho_{\rm min})\rangle$ & $\langle\log(\rho_{\rm max})\rangle$ & $\mathcal{M}_{\rm rms}$ & $\mathcal{M}_{\rm max}$ &Re & Re$_{\rm max}$ & Forcing type\\
  \hline
  Acmp & $4.0$ & $-4.29\pm 0.64$ & $1.51\pm 0.07$ & $3.24\pm 0.15$& $9.67\pm 0.69$ & $216\pm 10$ & $645\pm 46$ & compressive\\
  Bcmp & $8.0$ & $-5.28\pm 0.65$ & $1.77\pm 0.08$ & $5.33\pm 0.23$& $15.2\pm 1.80$ & $178\pm 7.7$ & $507\pm 60$ & compressive\\
  Ccmp & $12.0$ & $-6.03\pm 0.71$ & $1.91\pm 0.09$ & $7.01\pm 0.27$& $19.9\pm 2.56$ & $156\pm 6$ & $442\pm 57$ & compressive\\ 
    \hline
  Asol & $4.0$ & $-3.31\pm 0.52$ & $1.31\pm 0.06$ & $3.56\pm 0.10$& $9.57\pm 0.32$ & $237\pm 6.7$ & $638\pm 21$ & solenoidal \\
  Bsol & $8.0$ & $-4.42\pm 0.56$ & $1.55\pm 0.06$ & $5.62\pm 0.17$& $15.4\pm 1.22$ & $187\pm 5.7$ & $513\pm 41$ & solenoidal \\
  Csol & $12.0$ & $-5.29\pm 0.74$ & $1.70\pm 0.06$ & $7.45\pm 0.27$& $20.5\pm 2.73$ & $166\pm 6$ & $456\pm 61$ &solenoidal \\ 
  \hline
  \hline
  \end{tabular}
  \end{center}
  \end{table*}

            \begin{figure}
  \resizebox{\hsize}{!}{ 
    \includegraphics{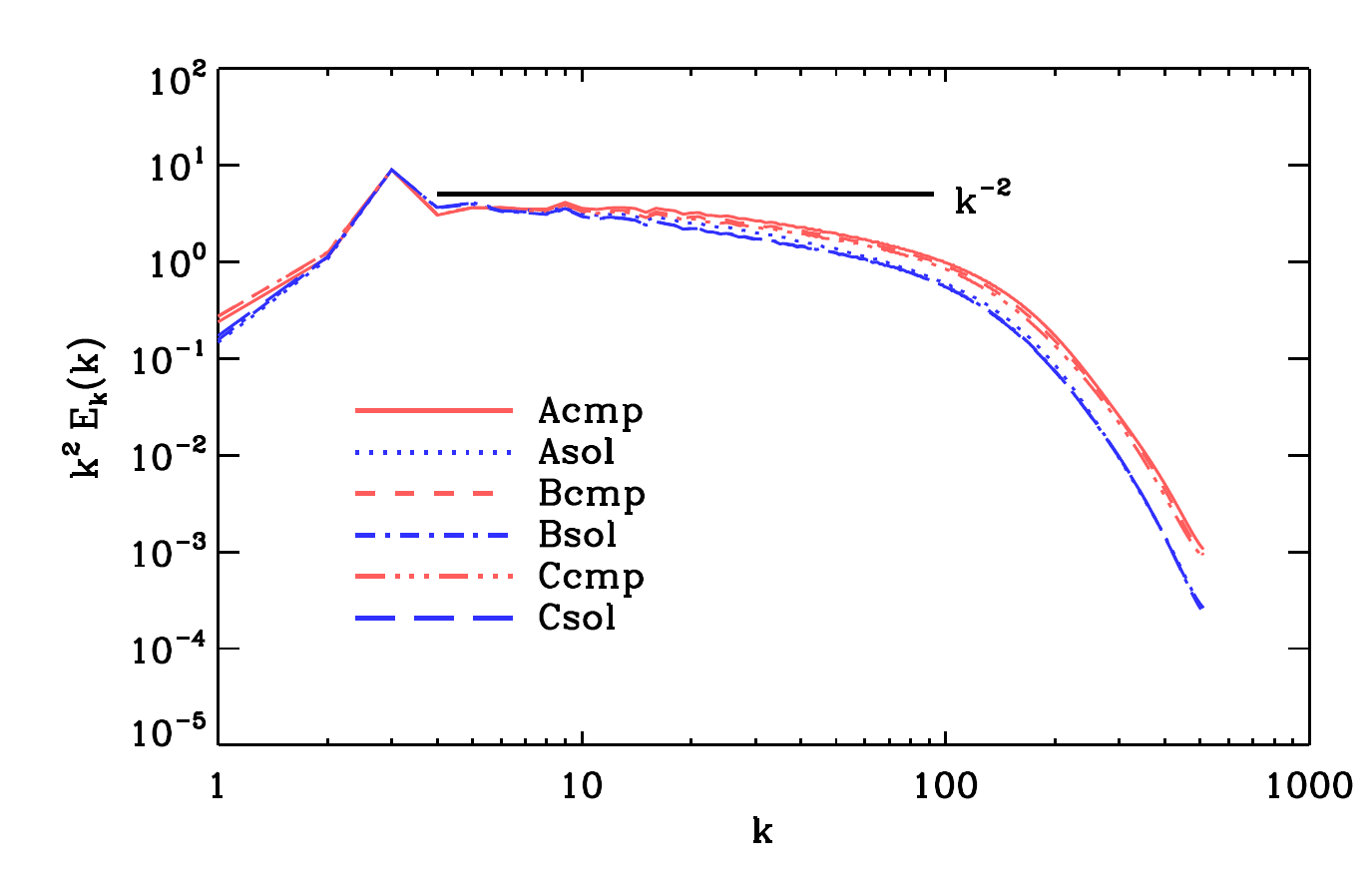}}
  \caption{\label{powerspectra} Kinetic-energy power spectra $E(k)$, multiplied by $k^{2}$, for the six simulations. The simulations with purely compressive forcing (Acmp, Bcmp and Ccmp) have spectra which are close to a Burgers spectrum ($E(k) \propto k^{-2}$) for wave numbers in the range $4\le k \le 20$. Simulations with purely solenoidal forcing (Asol, Bsol and Csol) show slightly steeper spectra in the same range.}
  \end{figure}

  \begin{figure*}
      \resizebox{0.9\hsize}{!}{
      \includegraphics{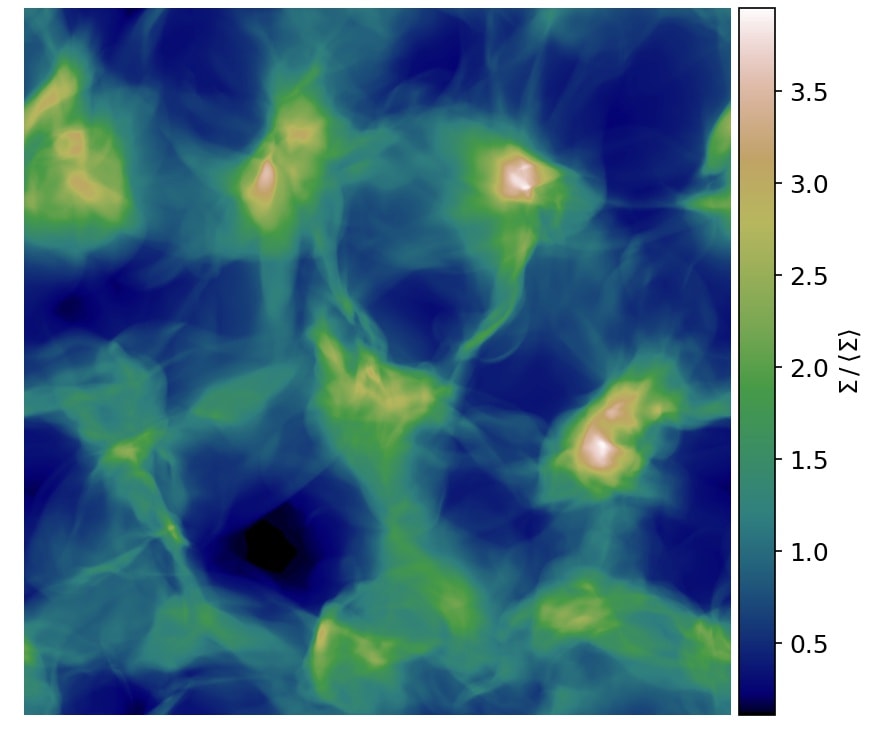}
      \includegraphics{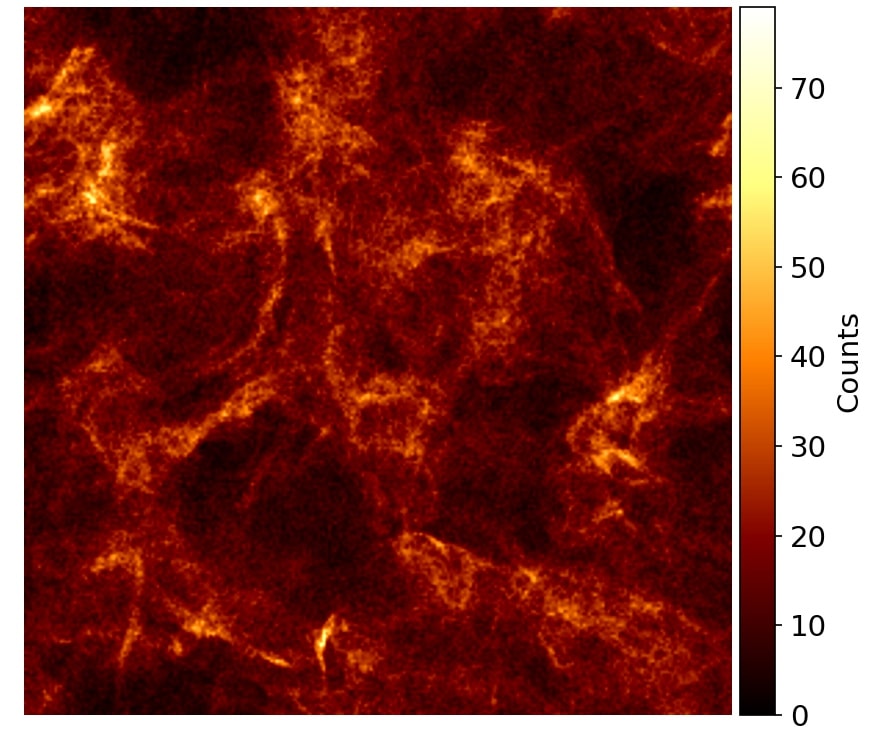}
      }
      \resizebox{0.9\hsize}{!}{
      \includegraphics{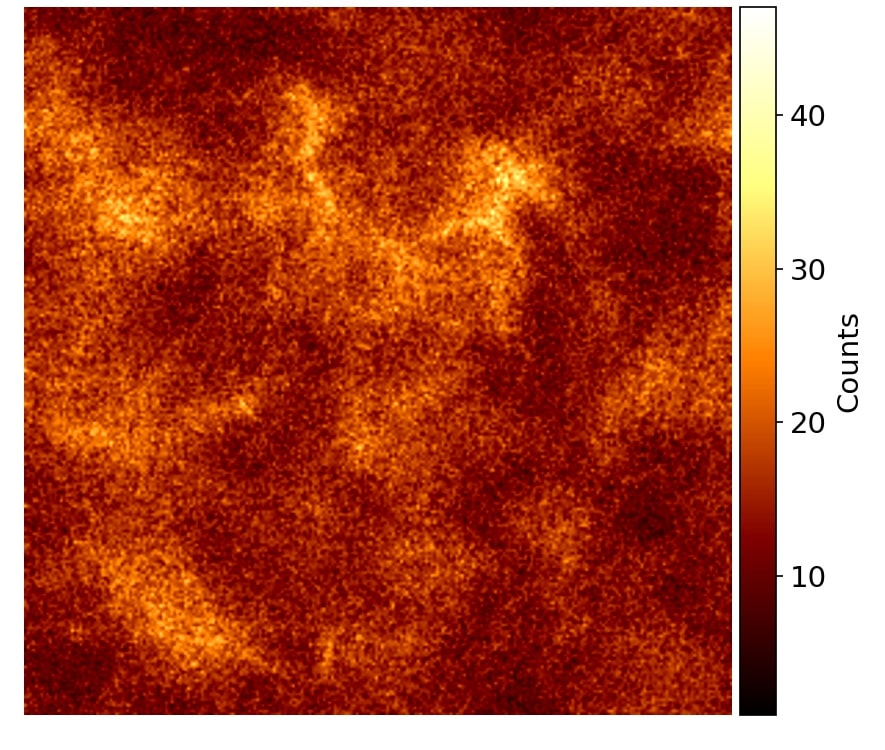}
      \includegraphics{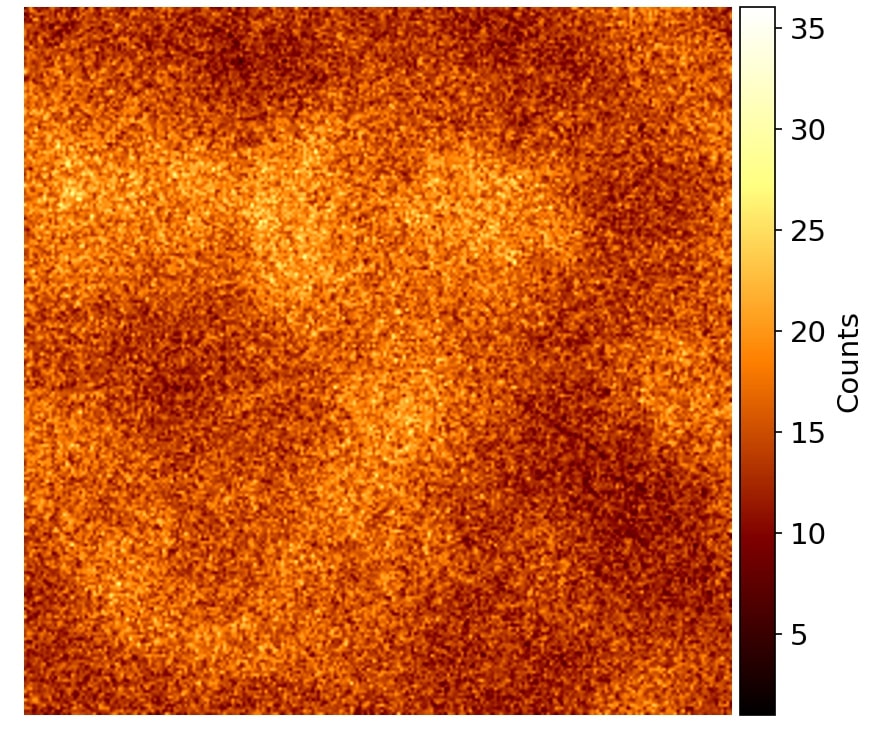}}
  \caption{\label{Ndust_Acmp} Column density of gas (upper left) and dust with $\alpha= 0.1$ (upper right), $\alpha= 0.5$ (lower left) and $\alpha= 1.5$ (lower right), for model Acmp (compressive forcing, $f=4.0$). The projections are calculated from snapshots taken at the end of the simulation.}
  \end{figure*}        
  \begin{figure*}
      \resizebox{0.90\hsize}{!}{
      \includegraphics{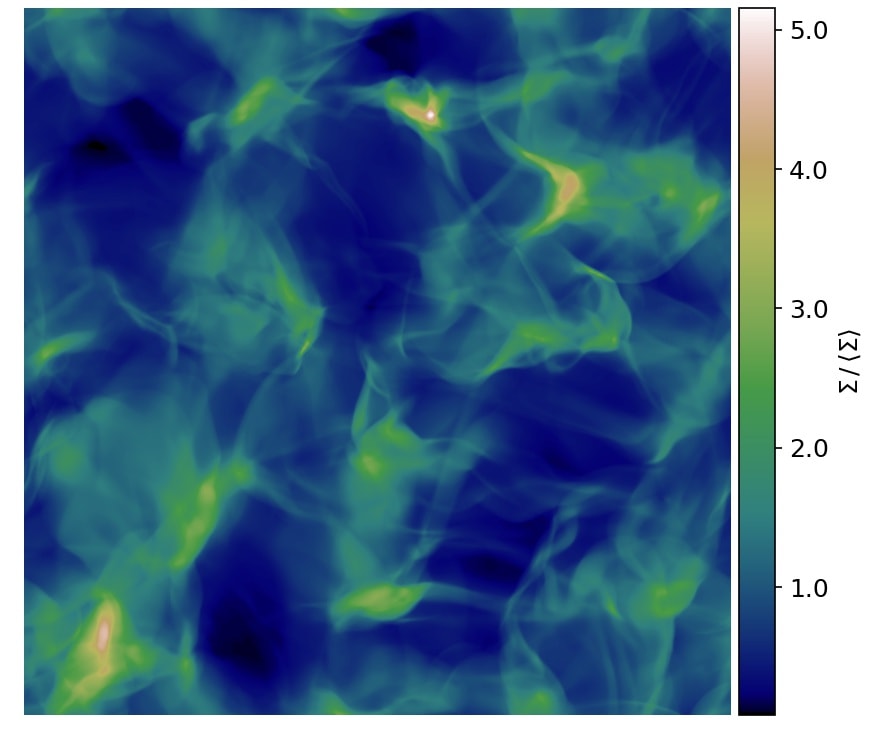}
      \includegraphics{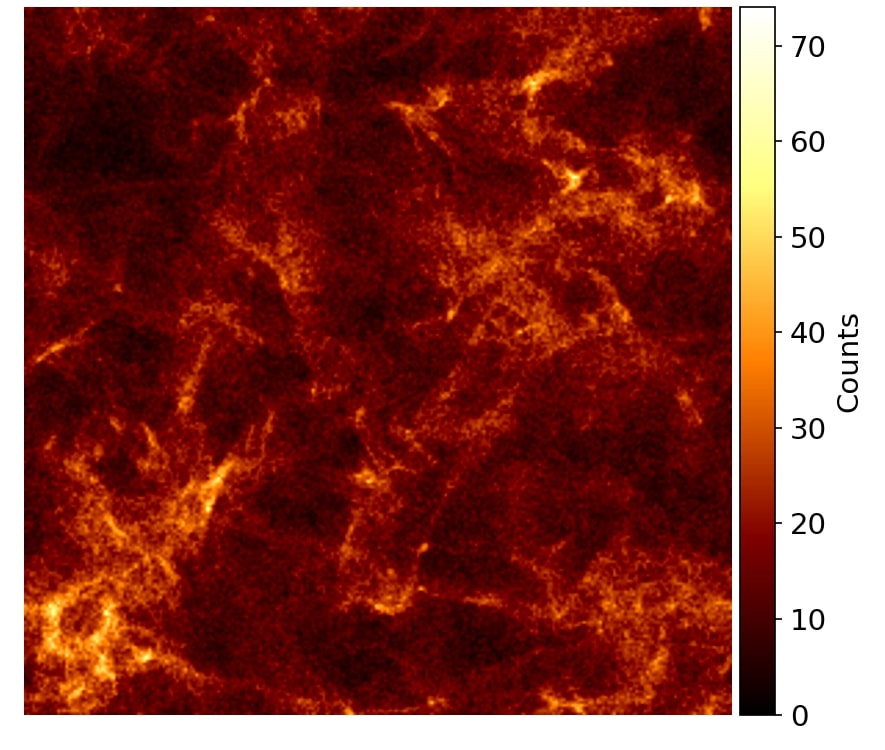}}
      \resizebox{0.90\hsize}{!}{
      \includegraphics{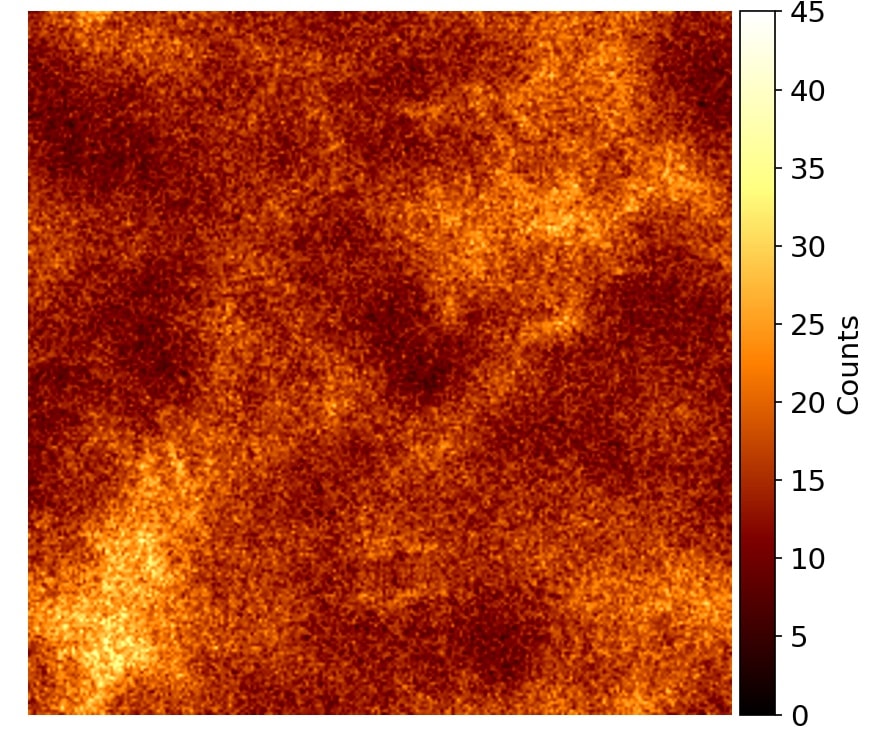}
      \includegraphics{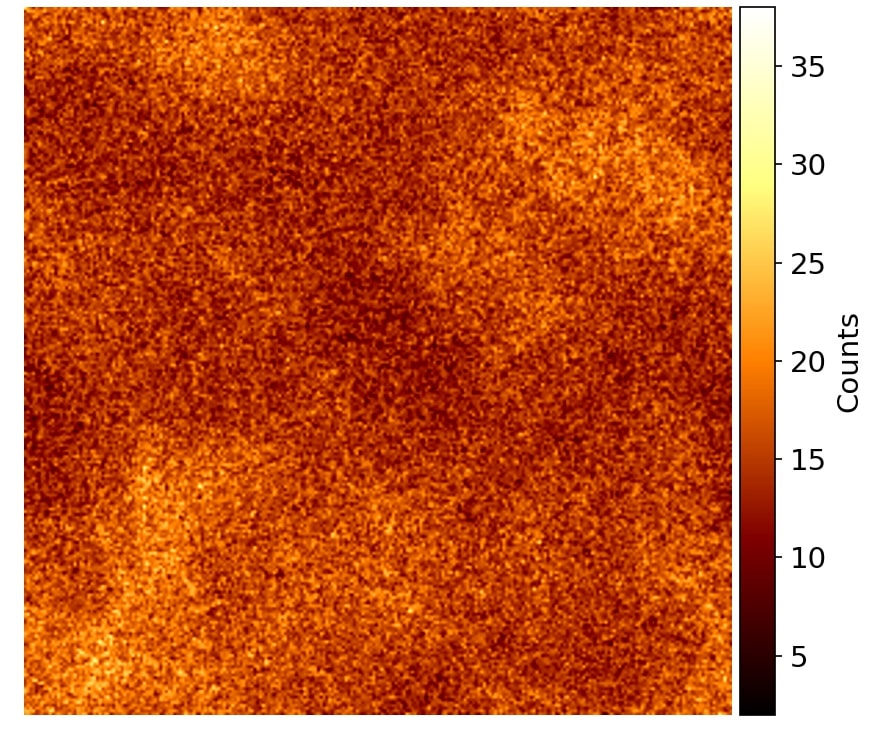}}
  \caption{\label{Ndust_Bcmp} Same as Fig. \ref{Ndust_Acmp}, but for model Bcmp (compressive forcing, $f=8.0$).}
  \end{figure*}   
  
  \begin{figure*}
      \resizebox{0.91\hsize}{!}{
      \hspace*{0.05 cm}
       \includegraphics{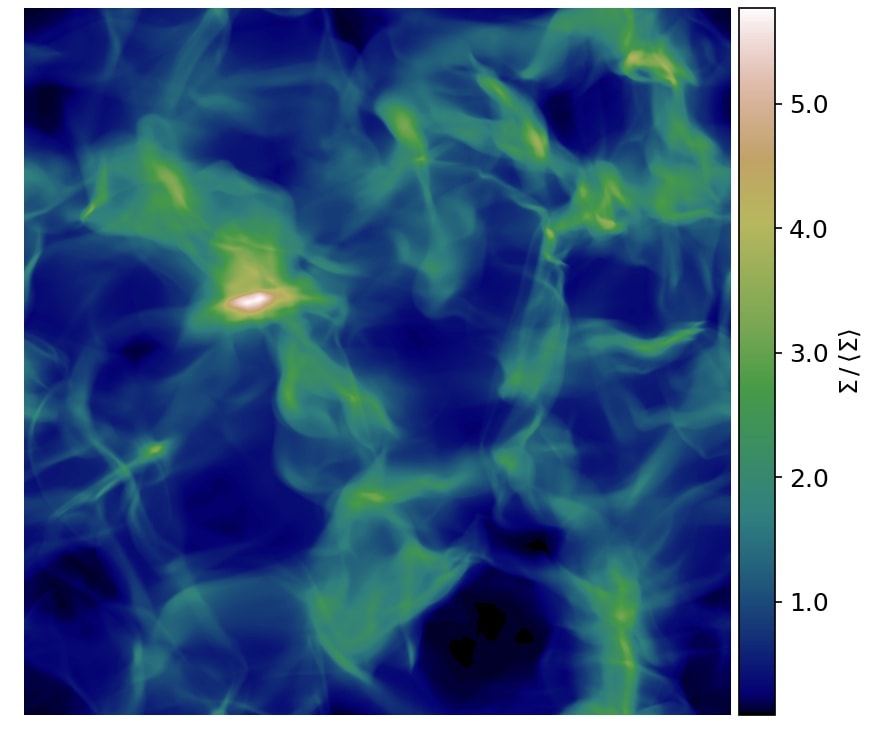}
       \hspace*{0.1 cm}
      \includegraphics{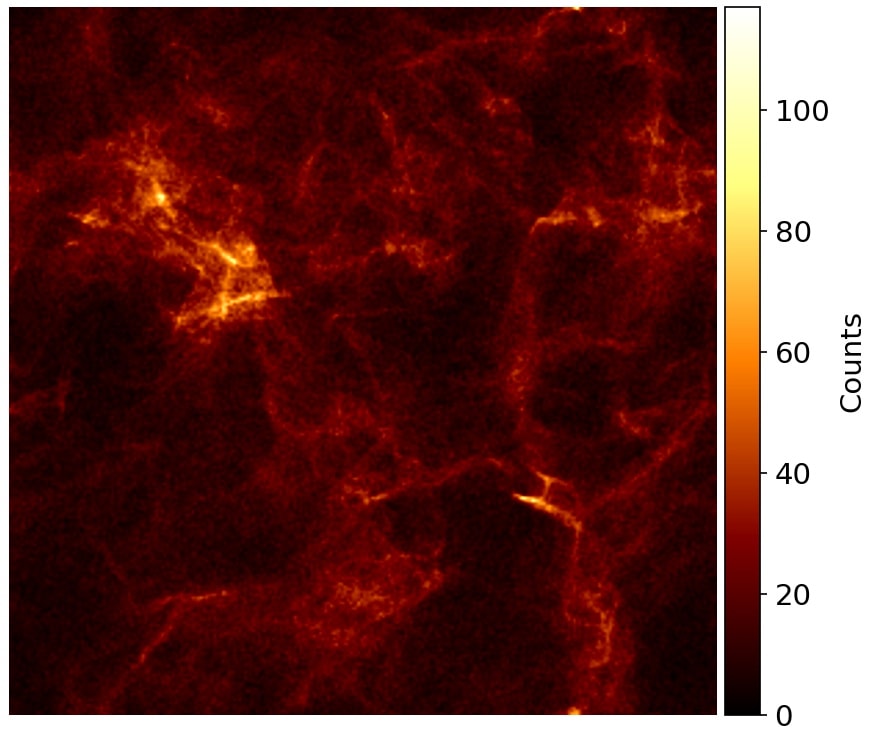}}
      \resizebox{0.90\hsize}{!}{
      \includegraphics{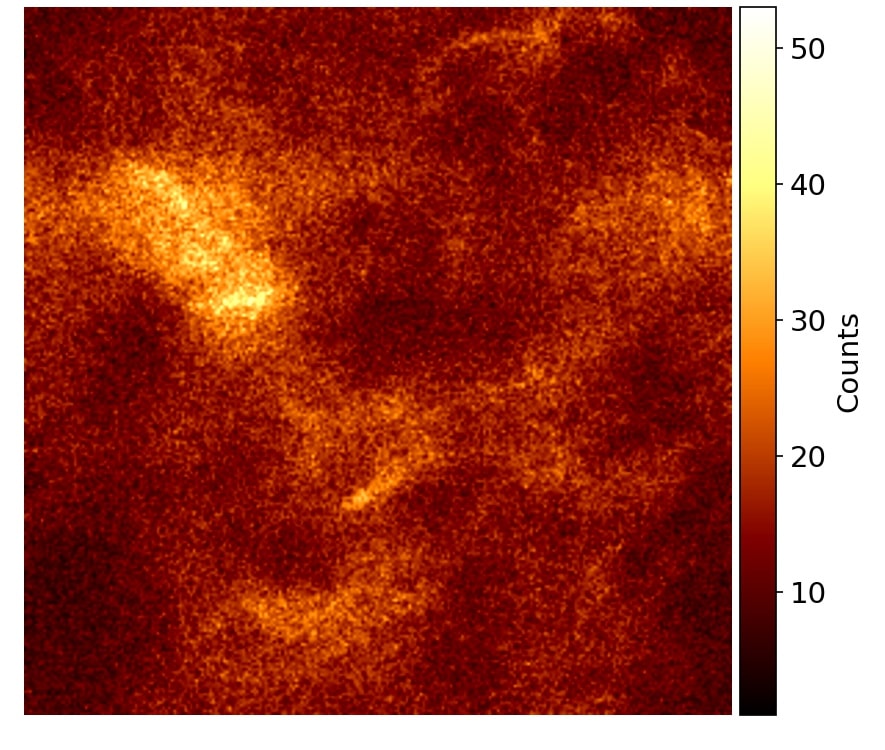}
      \includegraphics{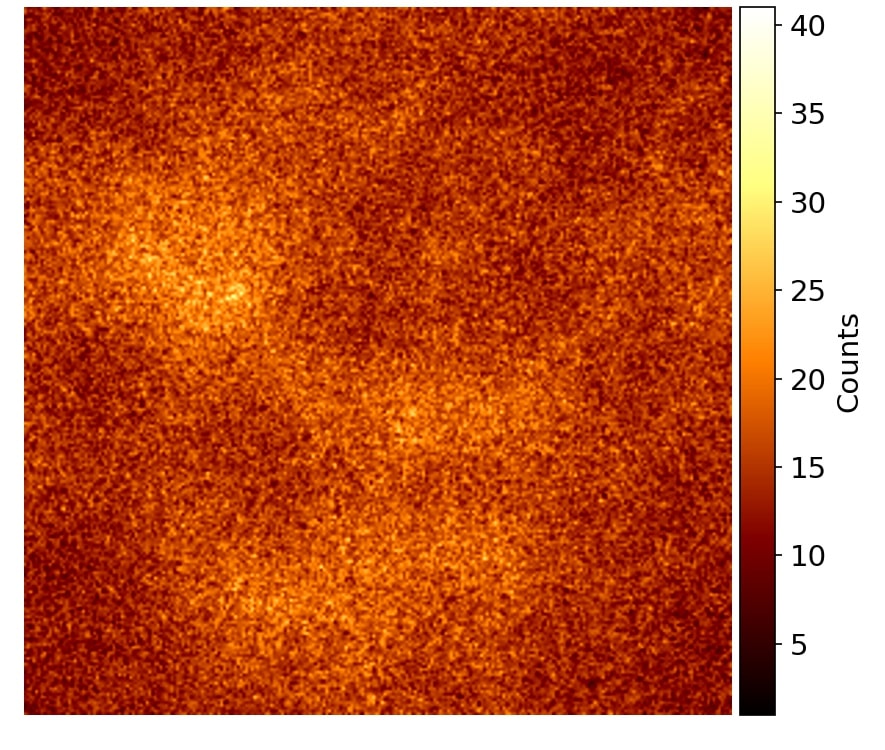}}
  \caption{\label{Ndust_Ccmp} Same as Fig. \ref{Ndust_Acmp}, but for model Ccmp (compressive forcing, $f=12.0$).}
  \end{figure*}

 \begin{figure*}
      \resizebox{0.90\hsize}{!}{
      \includegraphics{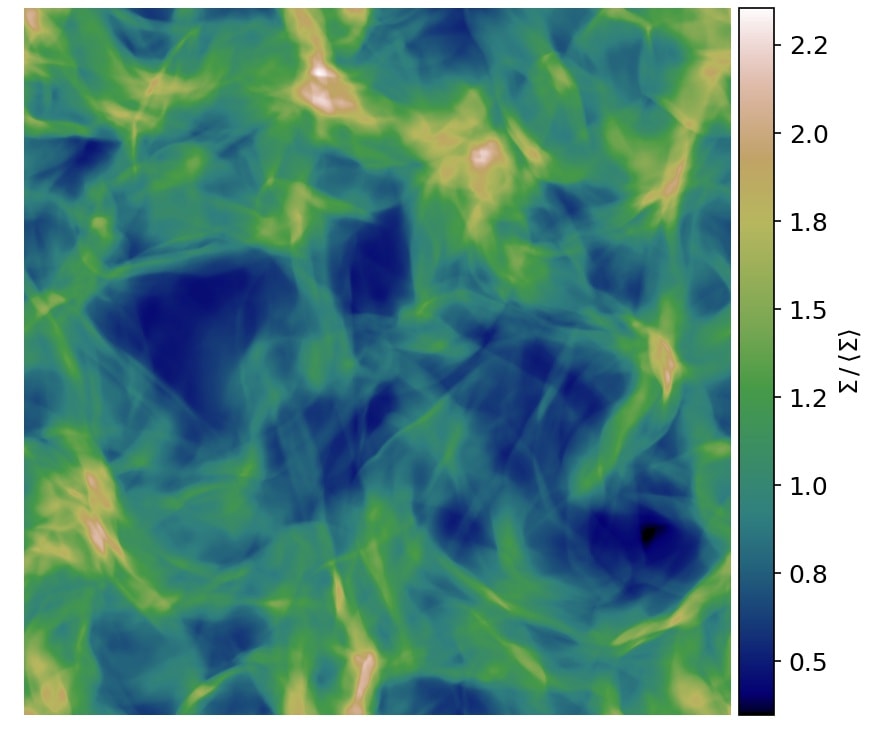}
      \includegraphics{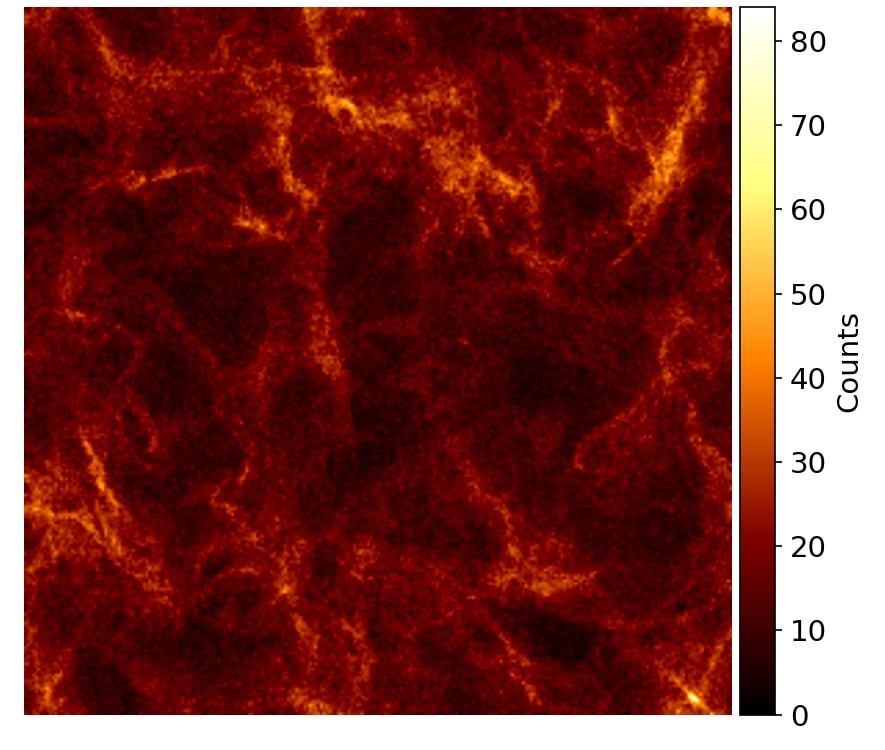}}
      \resizebox{0.90\hsize}{!}{
      \includegraphics{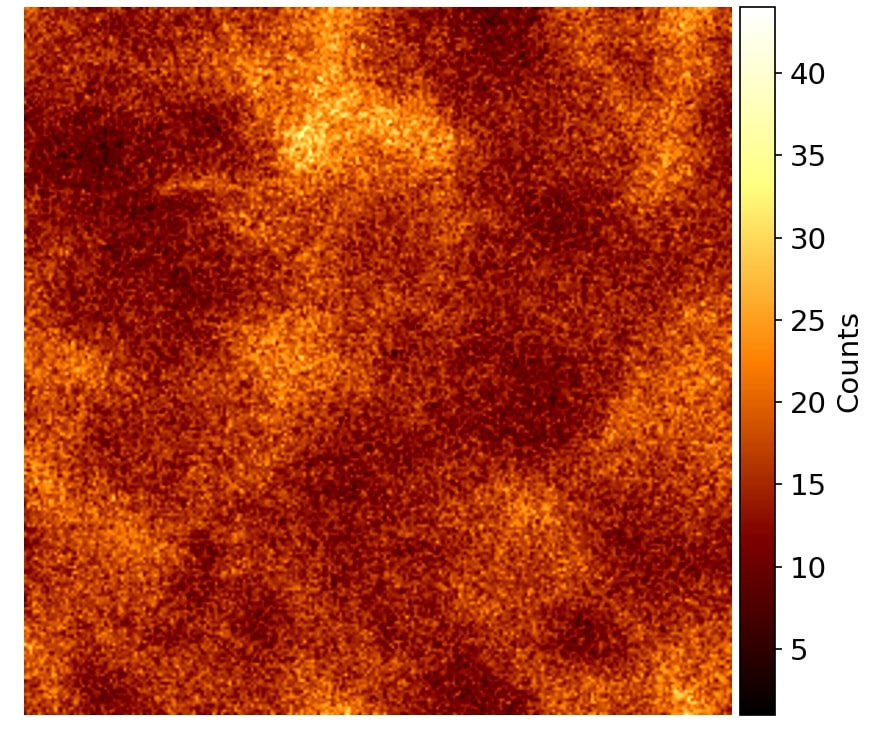}
      \includegraphics{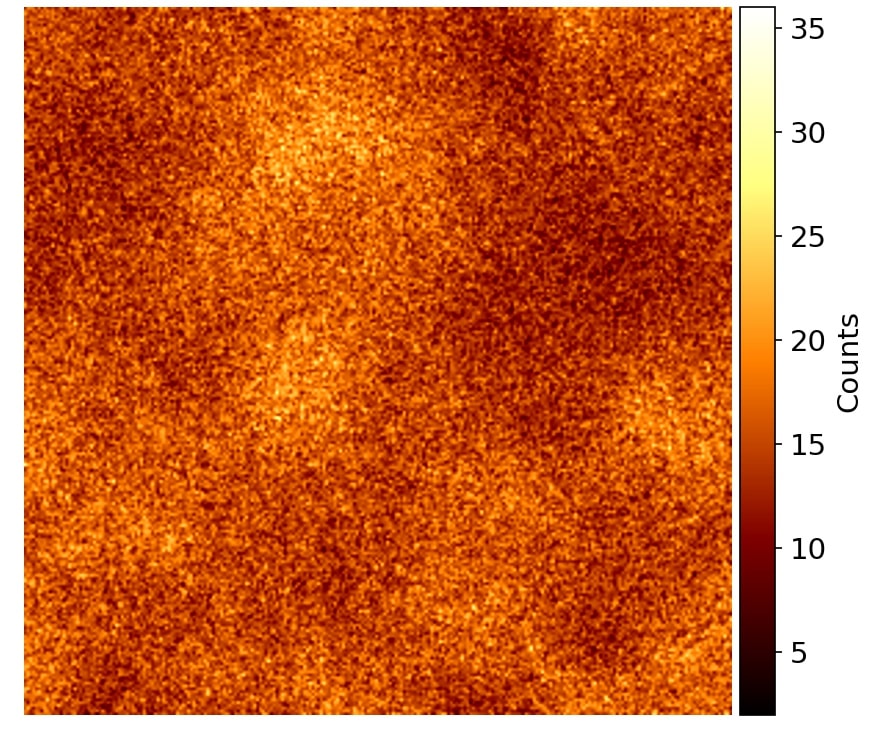}}
  \caption{\label{Ndust_Asol} Column density of gas (upper left) and dust with grain-size parameter $\alpha= 0.1$ (upper right), $\alpha= 0.5$ (lower left) and $\alpha= 1.5$ (lower right), for model Asol (solenoidal forcing, $f=4.0$). The projections are calculated from snapshots taken at the end of the simulation.}
  \end{figure*}
  
  \begin{figure*}
      \resizebox{0.90\hsize}{!}{
      \includegraphics{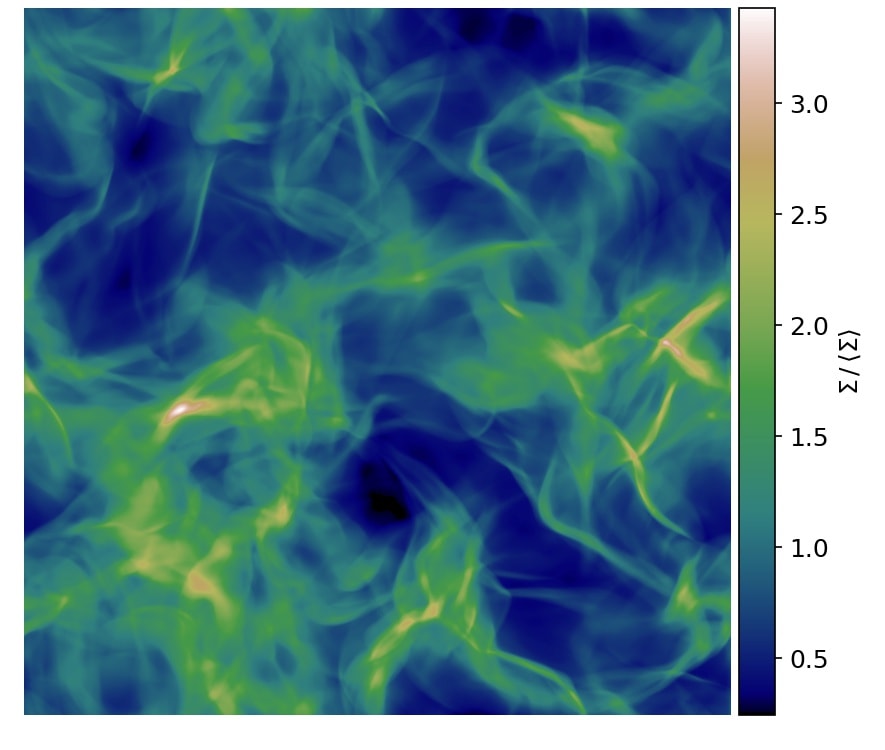}
      \includegraphics{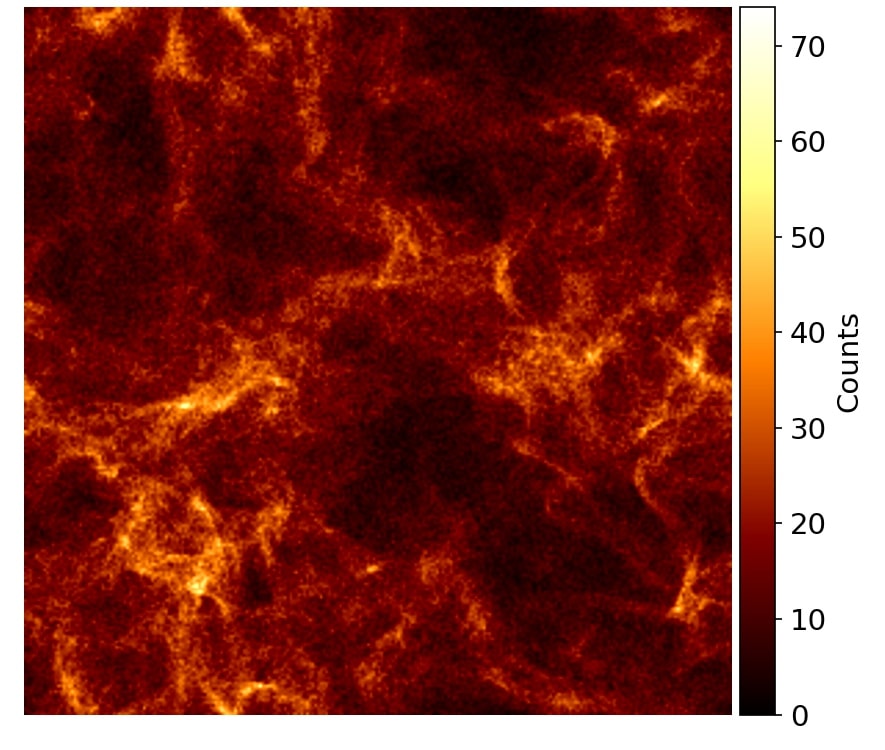}}
      \resizebox{0.90\hsize}{!}{
      \includegraphics{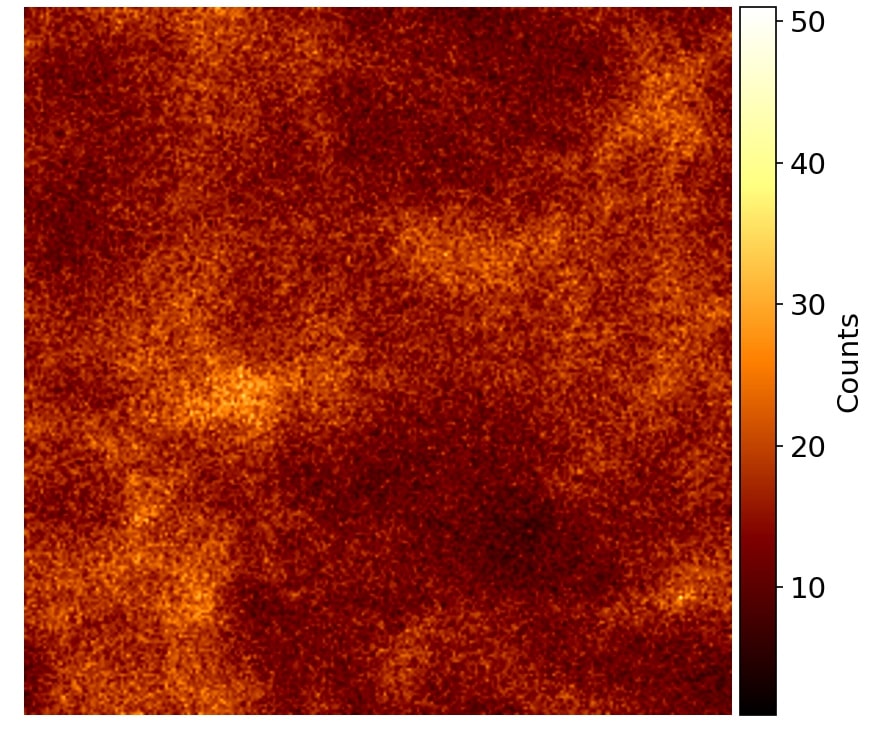}
      \includegraphics{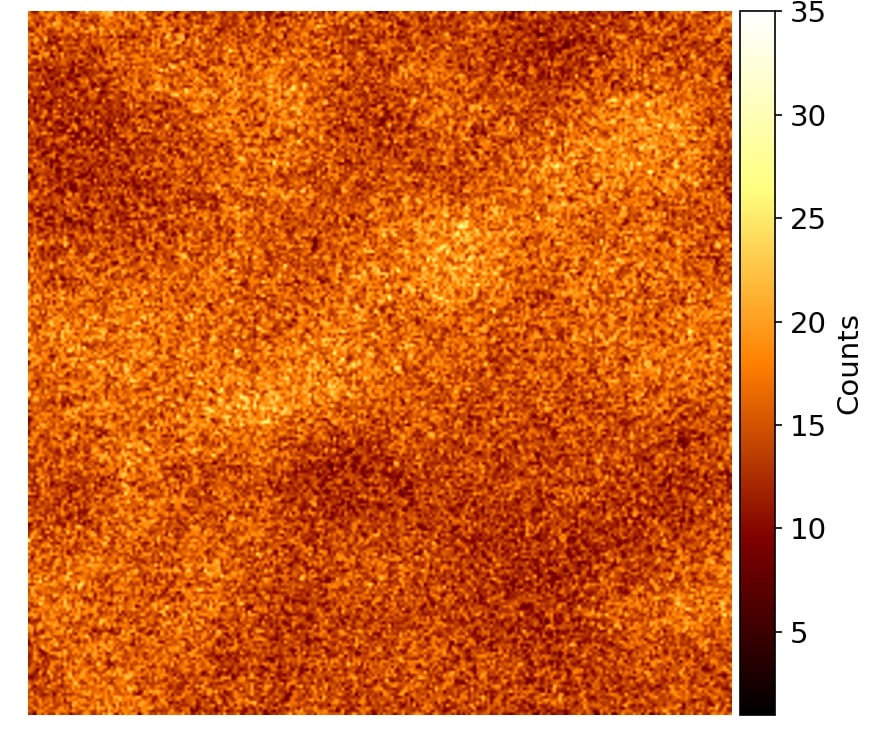}}
  \caption{\label{Ndust_Bsol} Same as Fig. \ref{Ndust_Asol}, but for model Bsol (solenoidal forcing, $f=8.0$).}
  \end{figure*}
  
  \begin{figure*}
      \resizebox{0.90\hsize}{!}{
      \includegraphics{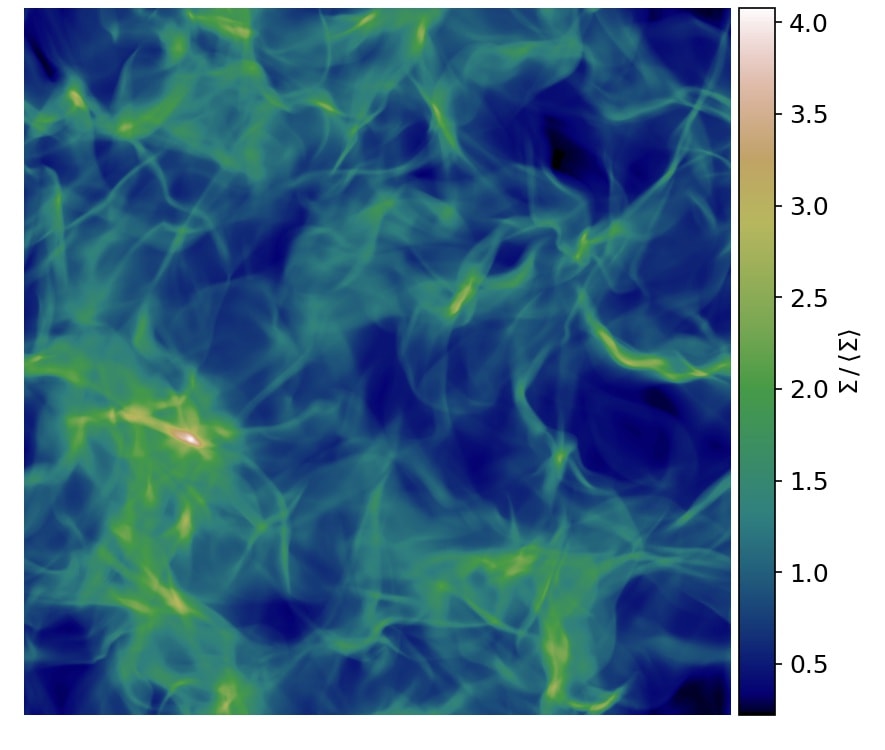}
      \includegraphics{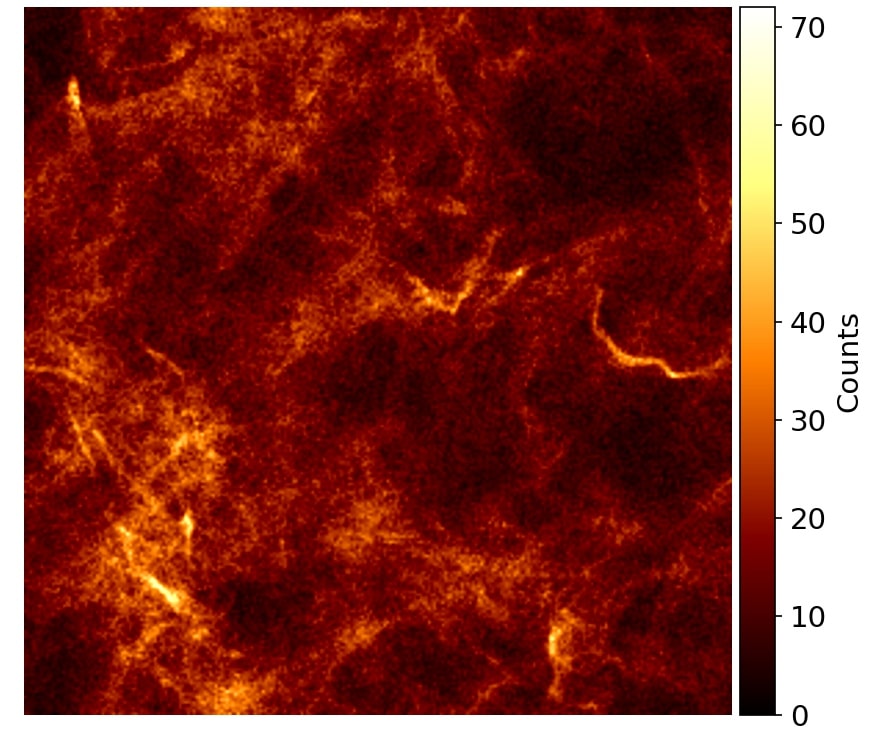}}
      \resizebox{0.90\hsize}{!}{
      \includegraphics{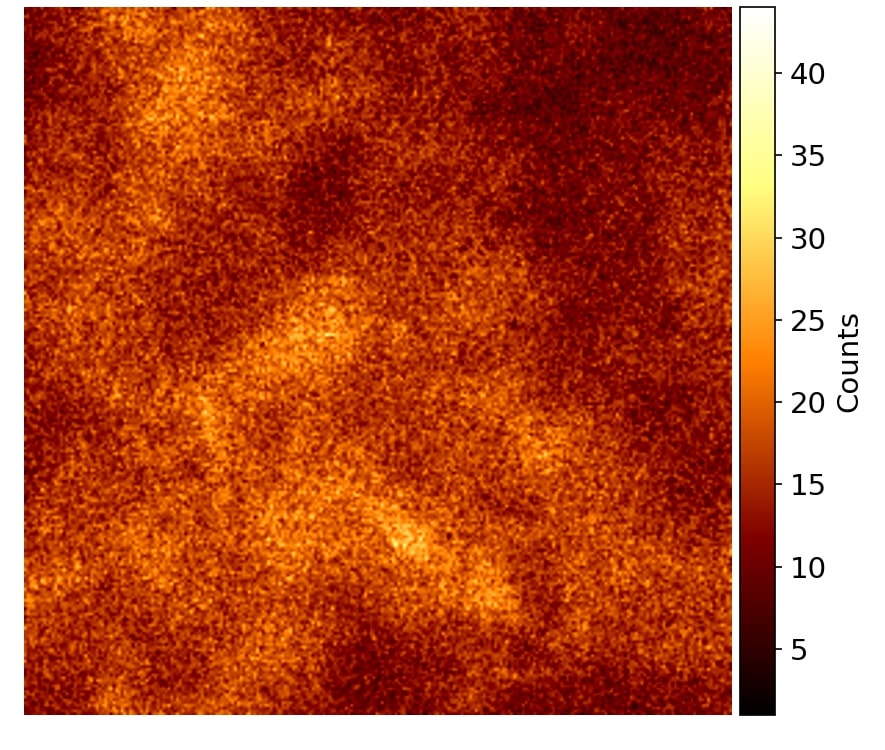}
      \includegraphics{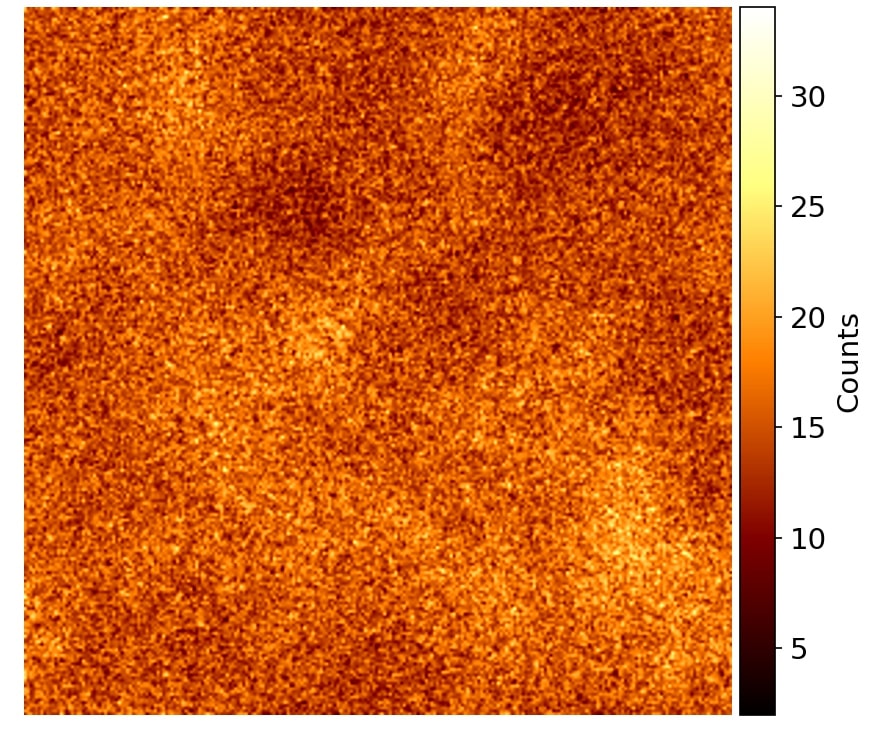}}
  \caption{\label{Ndust_Csol} Same as Fig. \ref{Ndust_Asol}, but for model Bsol (solenoidal forcing, $f=12.0$).}
  \end{figure*}
  
  \begin{figure*}
  \resizebox{\hsize}{!}{
   \includegraphics{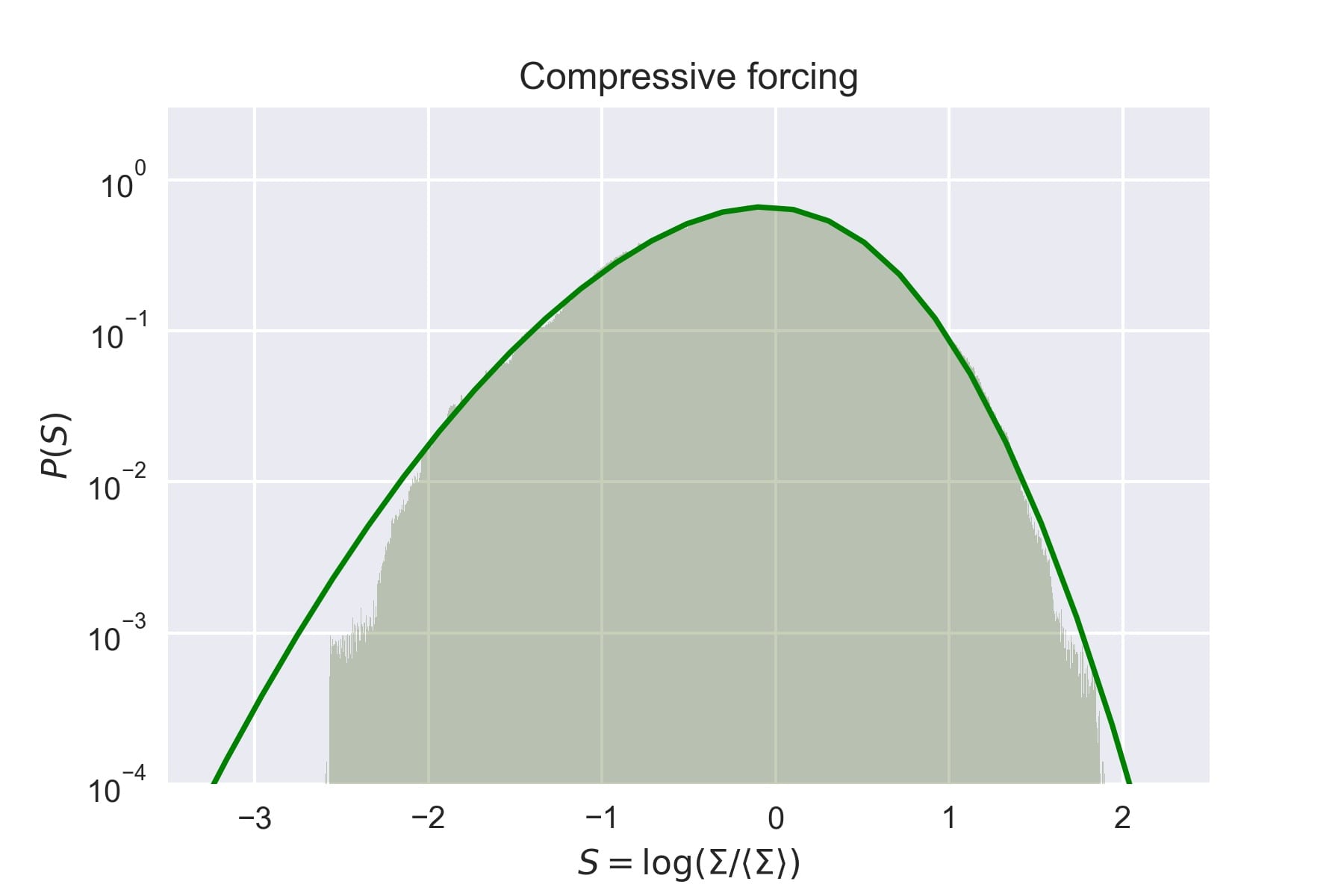}
   \includegraphics{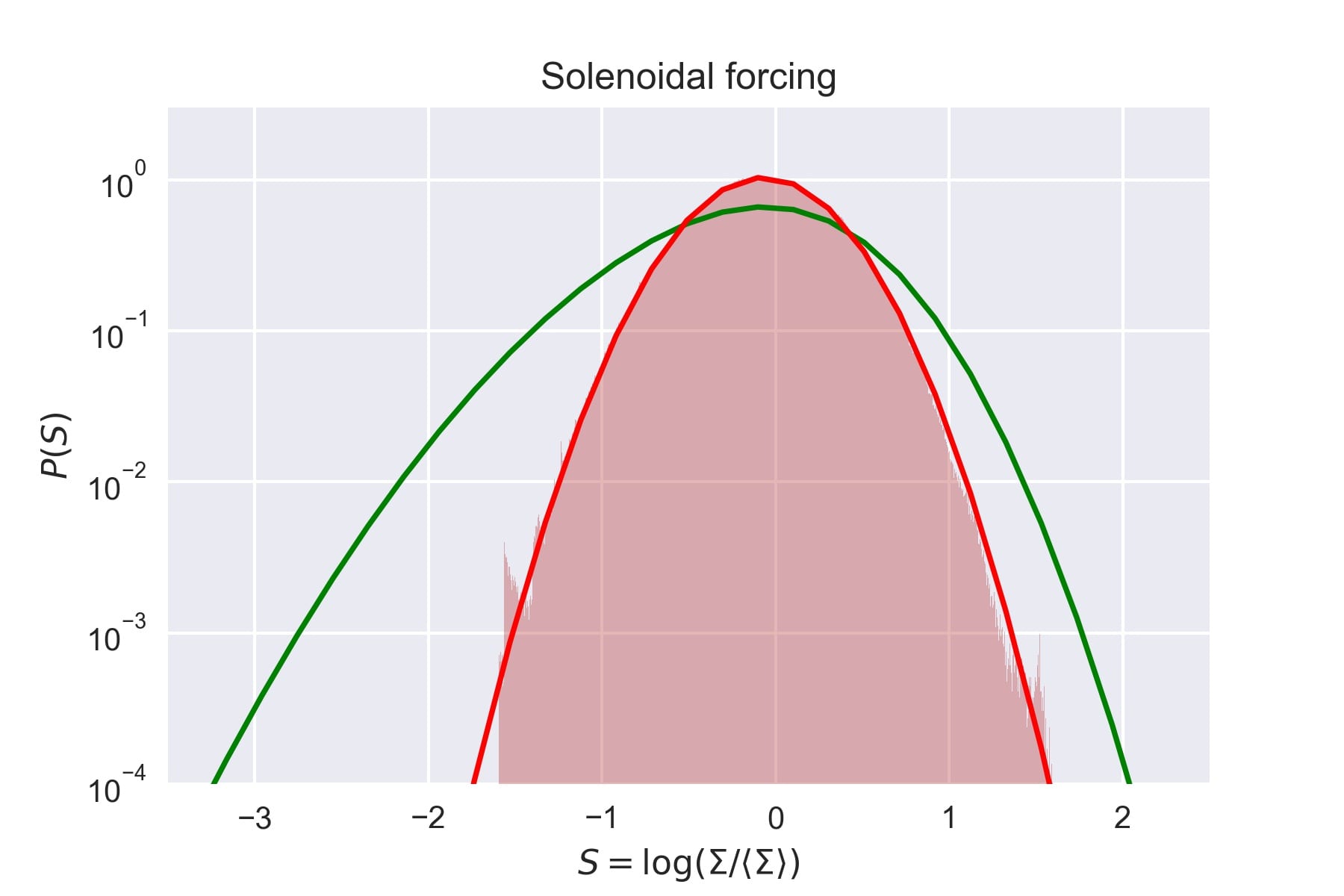}}
  \caption{\label{pdf_gas}  Probability density functions (PDFs) for the projected gas density in case of compressive forcing (left) and solenoidal forcing (right) with $f = 4.0$. The green line shows a fit of a skewed lognormal PDF to the measured PDF of the simulation with compressive forcing. The red line (right panel) is a regular lognormal fit to the measured PDF of the simulation with solenoidal forcing. }
  \end{figure*}
  
 \begin{figure}
  \resizebox{\hsize}{!}{ 
    \includegraphics{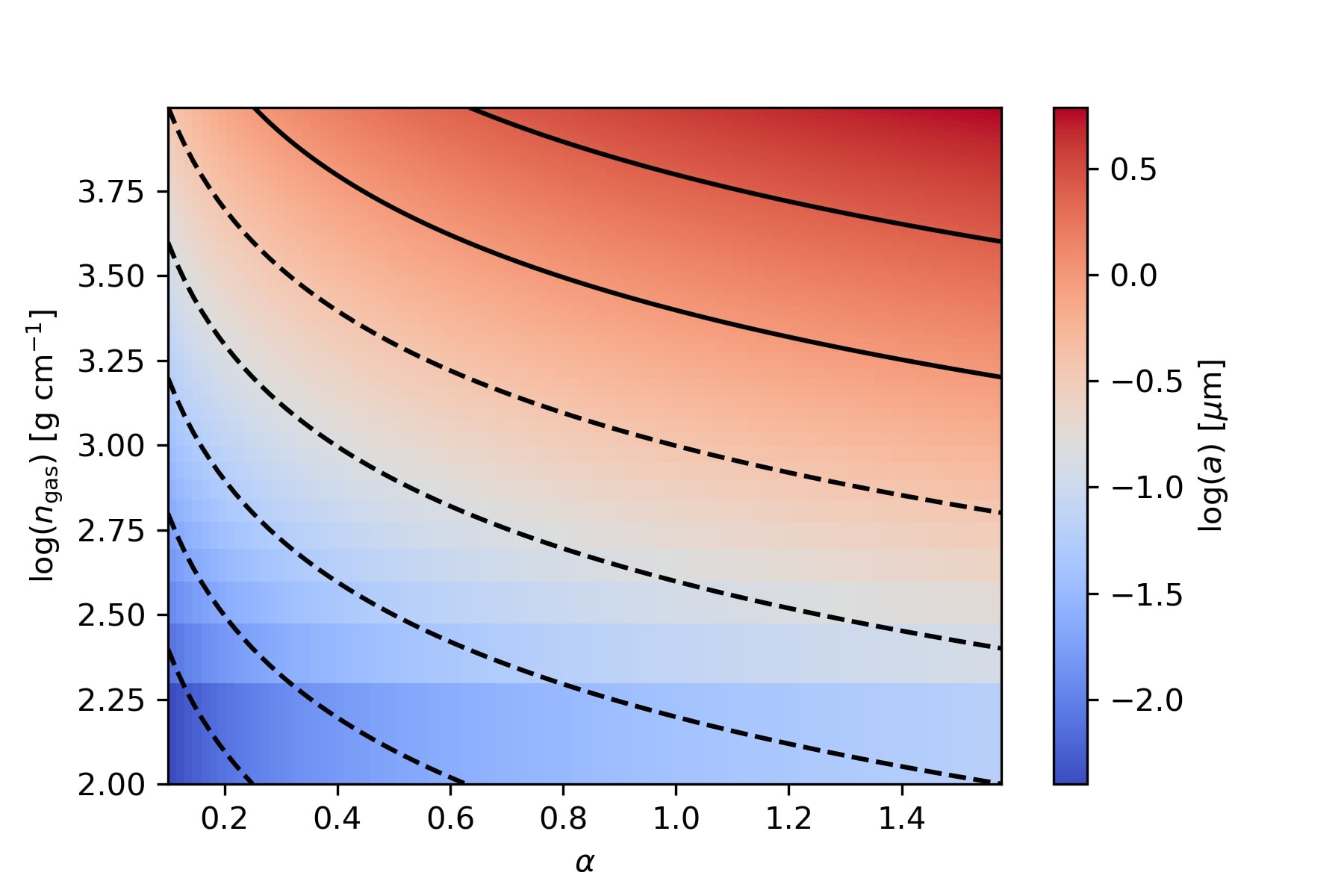}}
  \caption{\label{agrain} Grain sizes in physical units as a function of the grain-size parameter $\alpha$ evaluated at different mean gas densities, assuming a physical size of the simulation box $L_x=L_y=L_z=0.1$~pc. }
  \end{figure}
  
          \begin{figure*}
  \resizebox{\hsize}{!}{
   \includegraphics{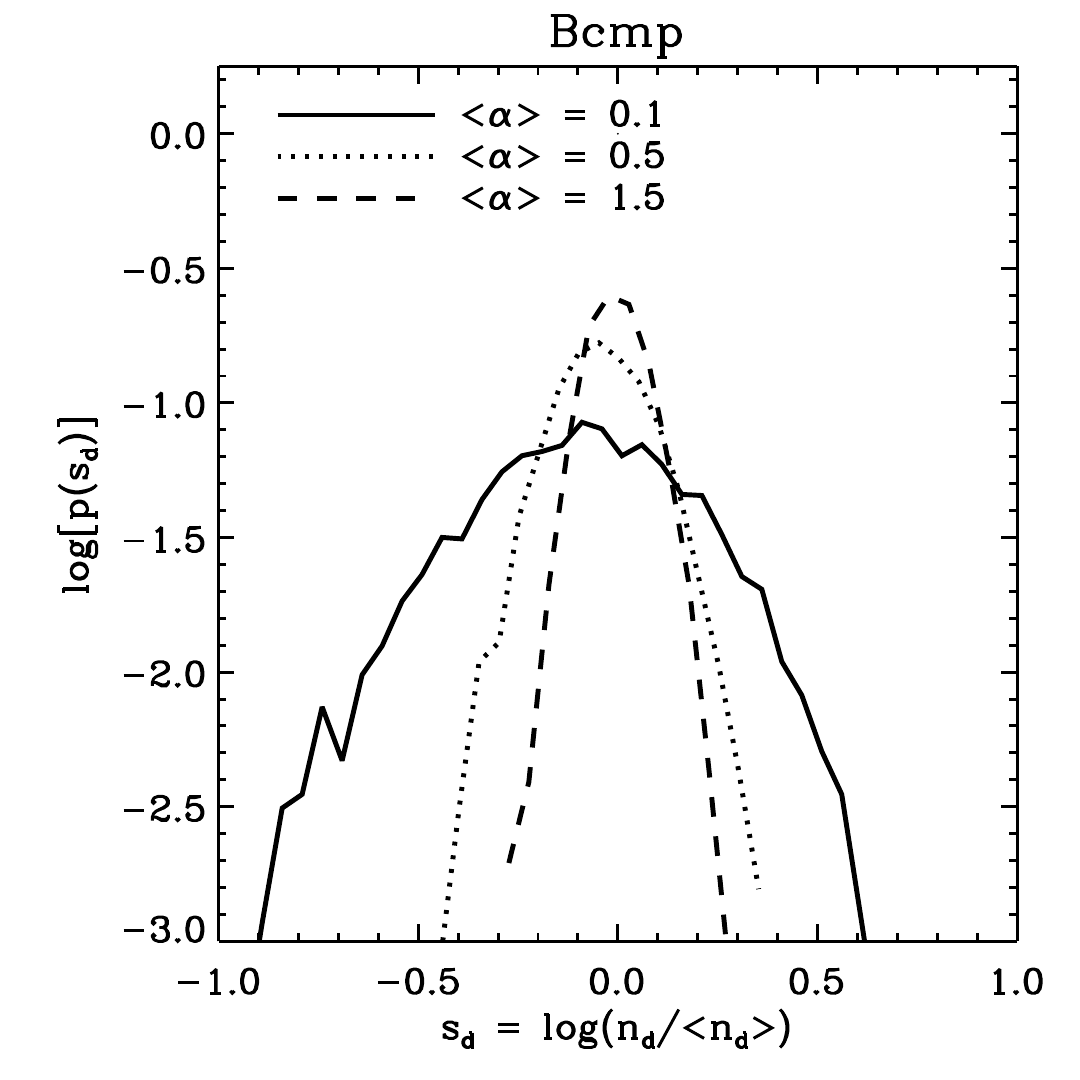}
   \includegraphics{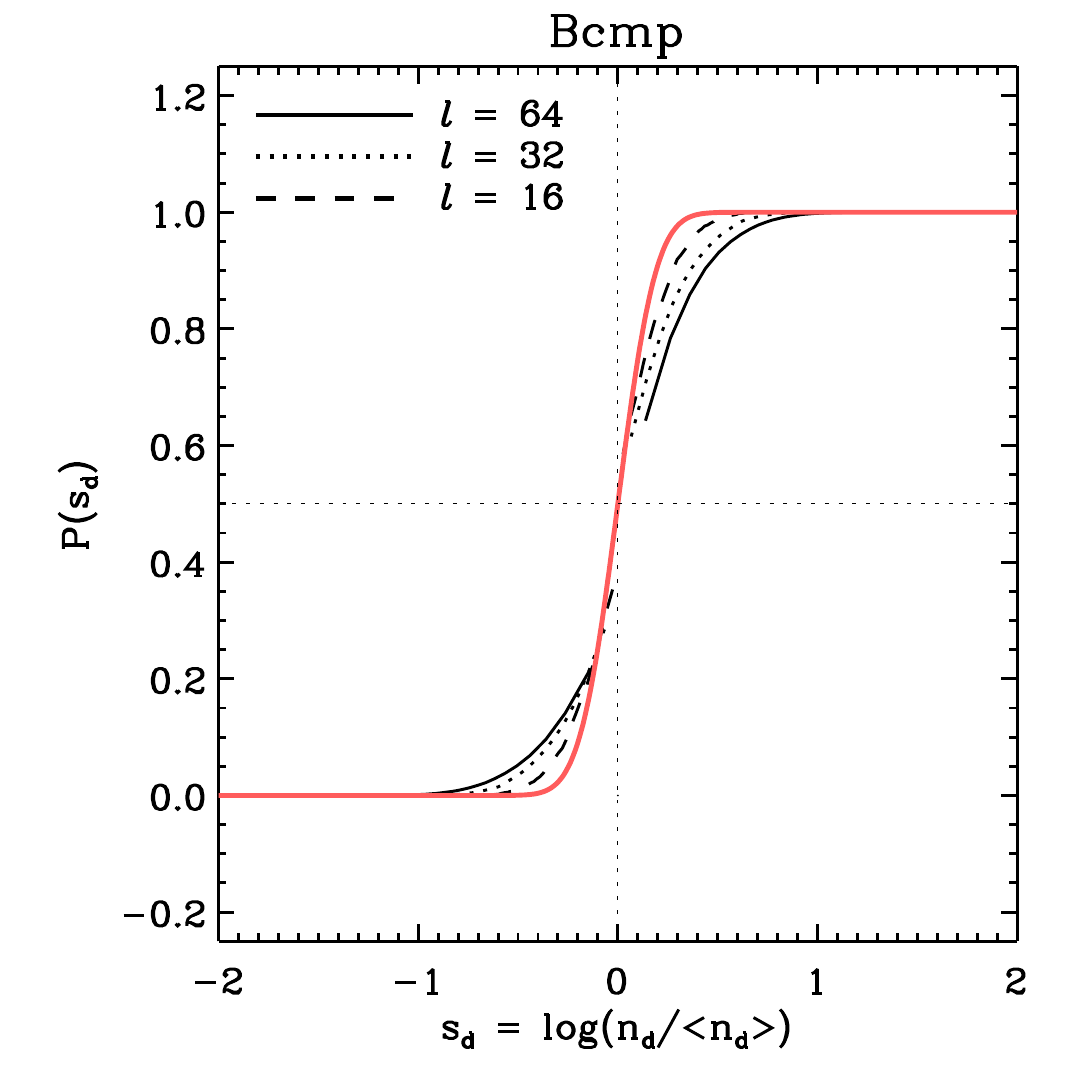}
   \includegraphics{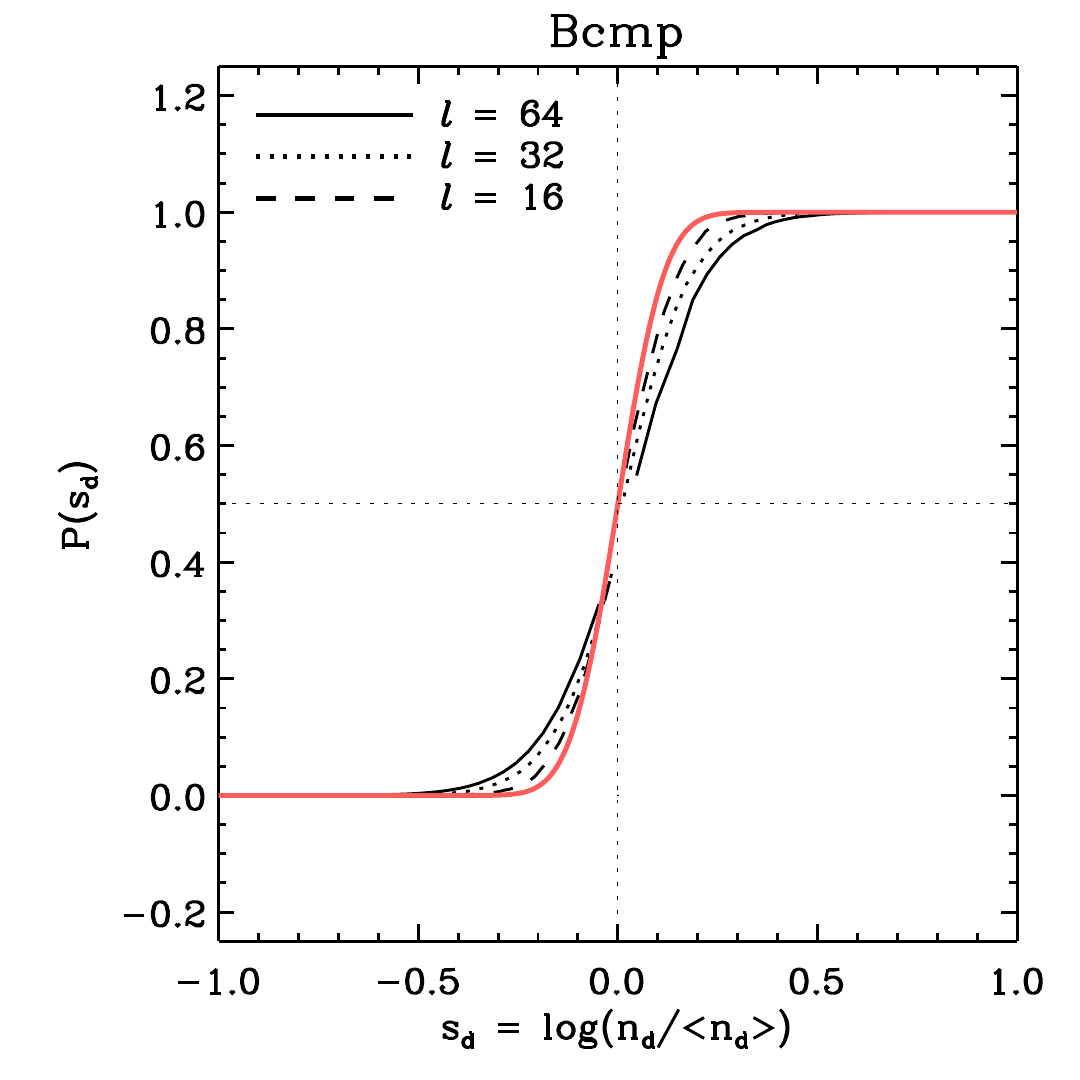}
   \includegraphics{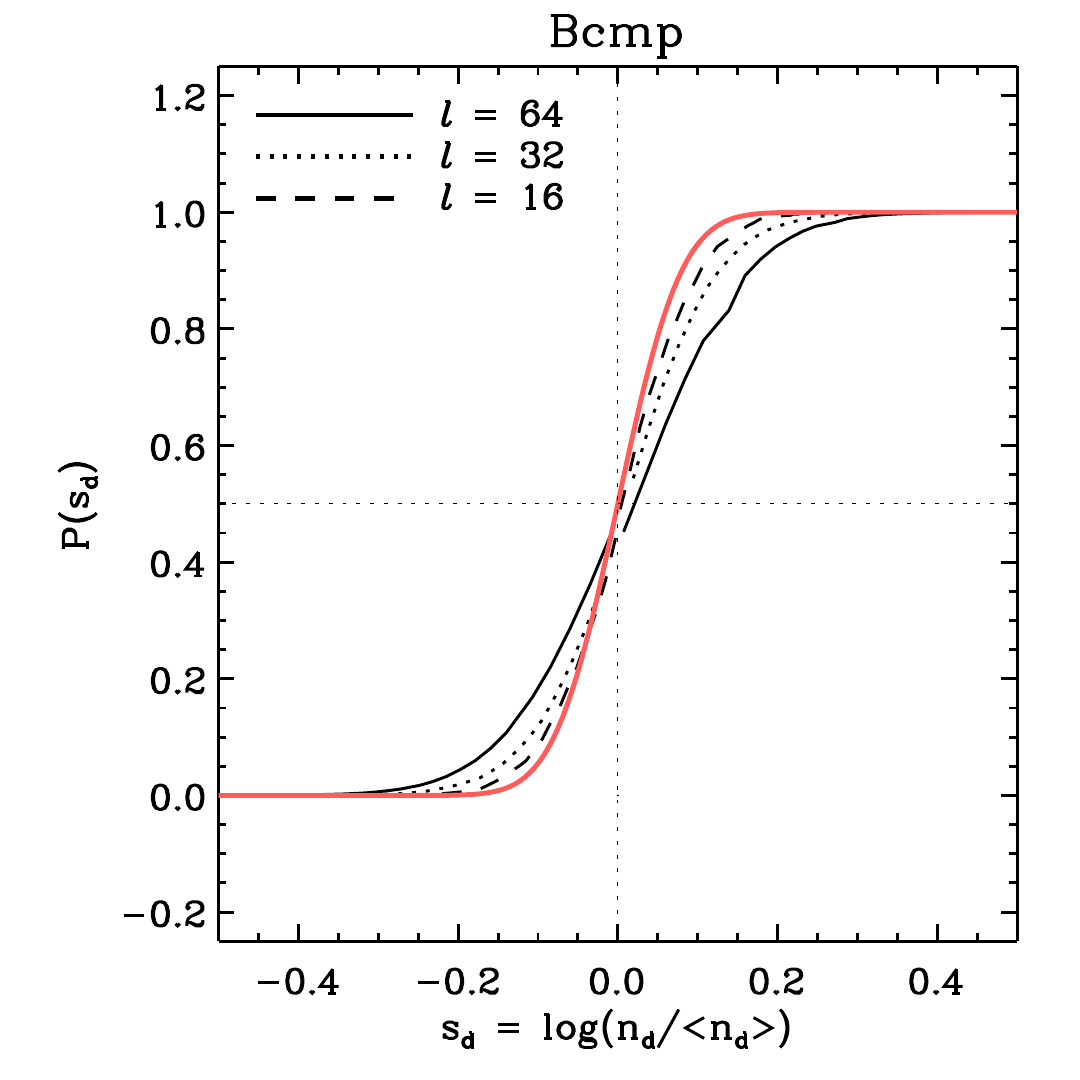}}
  \resizebox{\hsize}{!}{
   \includegraphics{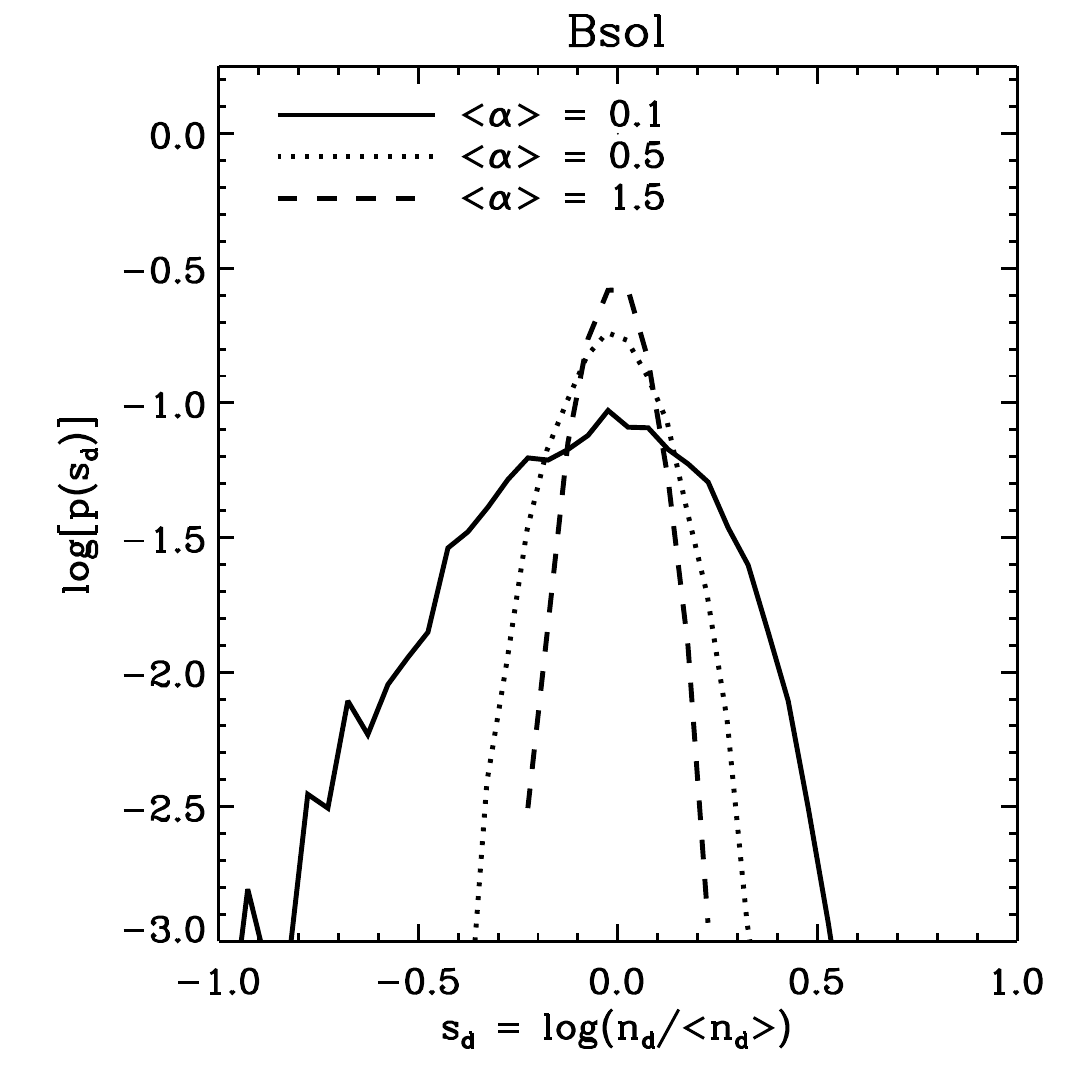}
   \includegraphics{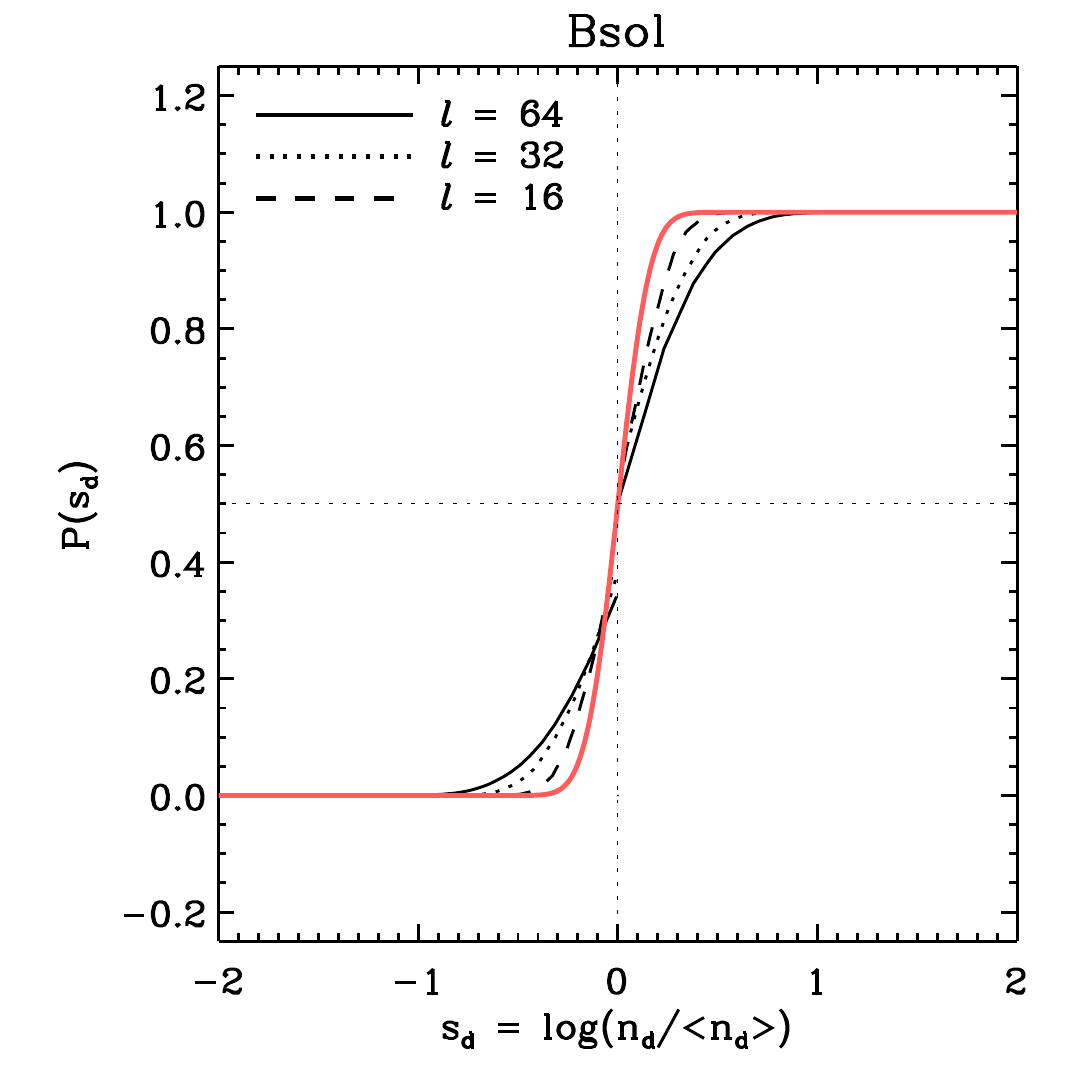}
   \includegraphics{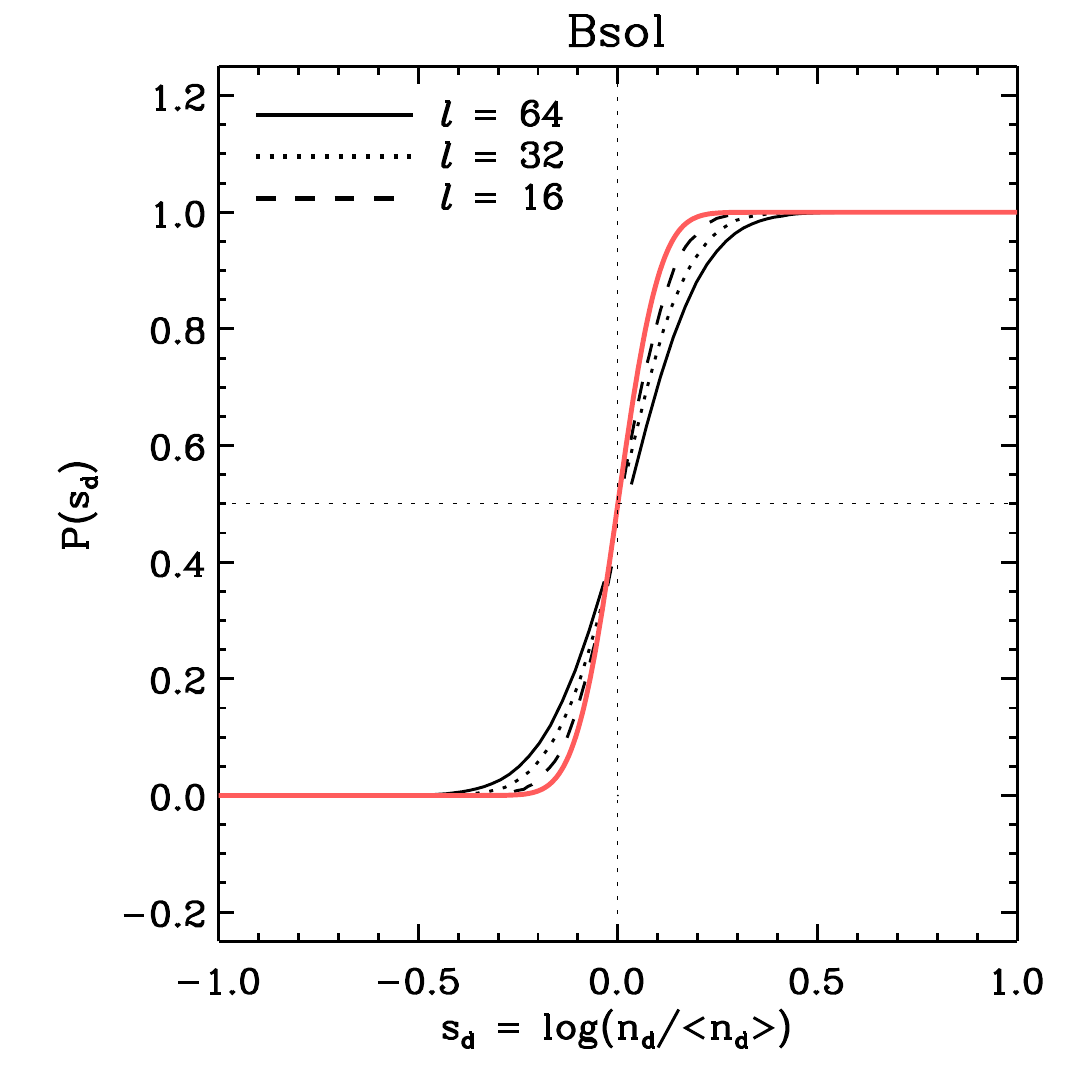}
   \includegraphics{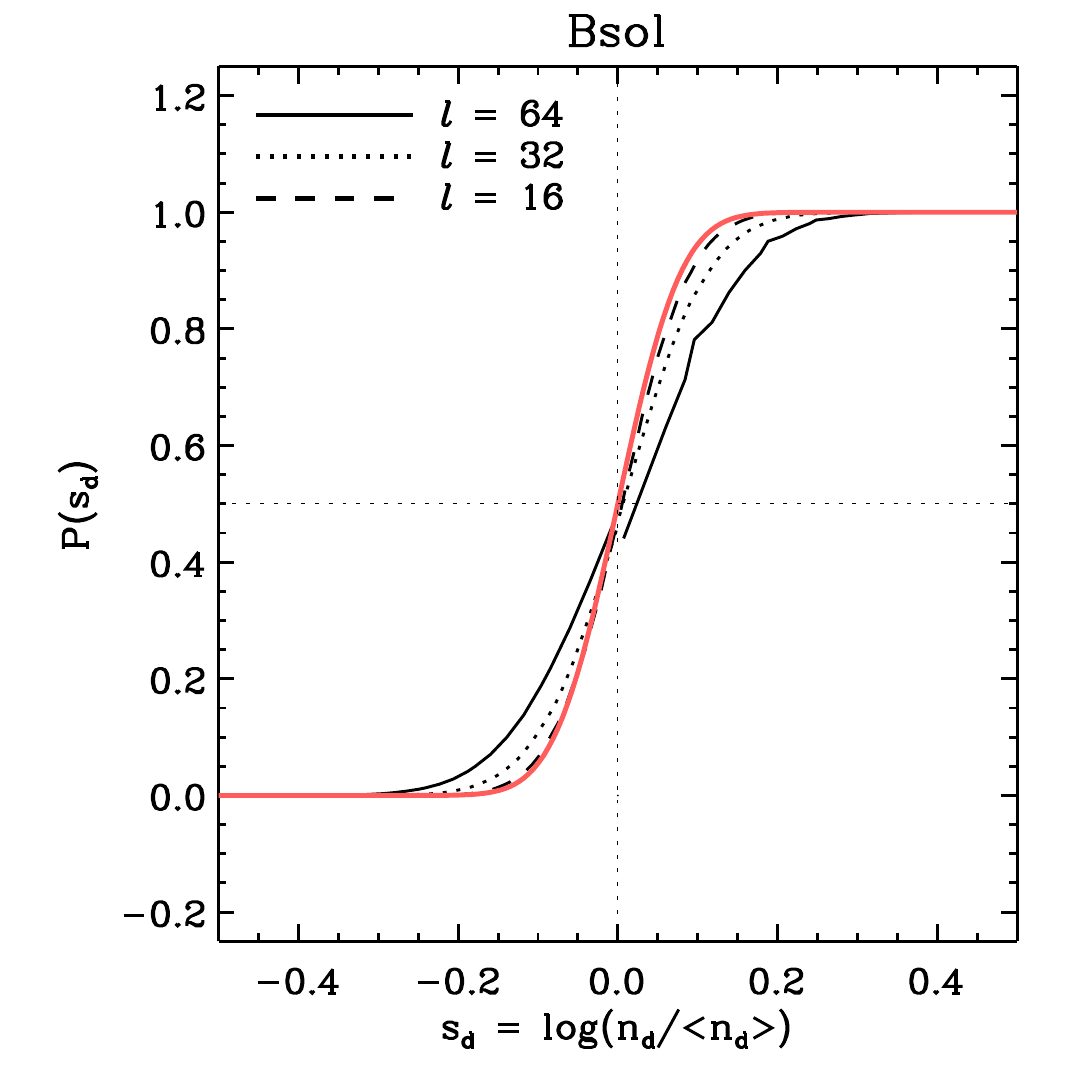}}
  \caption{\label{dustPDFs} Upper panels: PDFs for the number density of dust for three selected grain sizes ($\alpha = 0.1, 0.5, 1.5$) and compressive forcing with a forcing factor $f=8.0$ (left), together with cumulative density functions (CDFs) for $\alpha = 0.09$ (middle) and$\alpha = 10.0$ (right), which show that the PDFs for large and small grains indeed approaches a lognormal form (black lines in the left panel represent lognormal distributions). Lower panels: same as the upper panels, but for the case of purely solenoidal forcing.}
  \end{figure*}

  \begin{figure}
  \resizebox{\hsize}{!}{
   \includegraphics{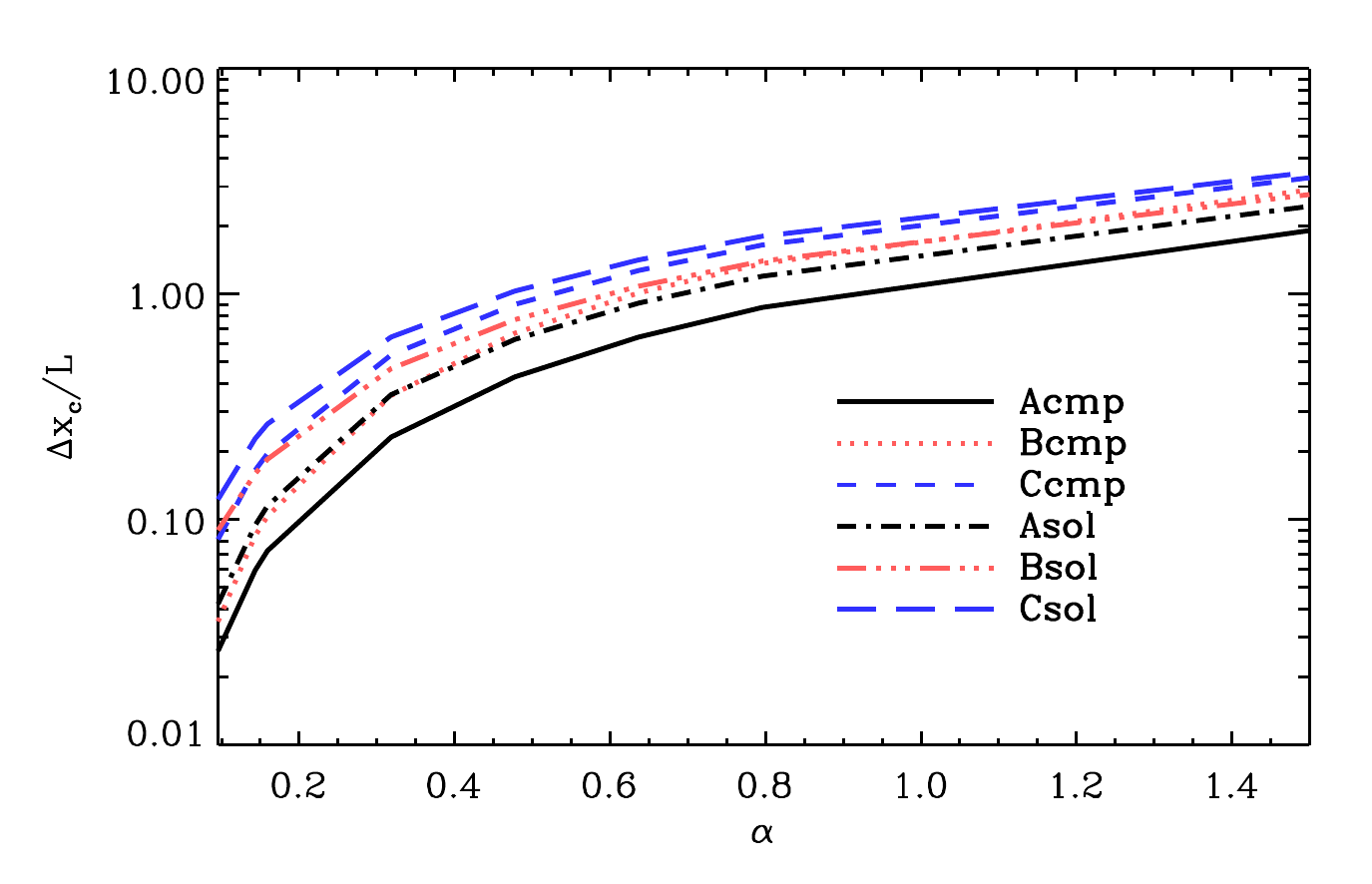}}
  \caption{\label{clength} {Coupling length (relative to the size of the simulation box) as a function of the grain-size parameter $\alpha$ for all six simulations. The simulations are limited to 10 different $\alpha$ values, but the result can easily be interpolated to any $\alpha$. For $\alpha$-values larger than 0.5 -- 1.0 (the range where the grains decouple from the flow) there is a correspondence with the size of the simulation box, indicating that the dust has lost its ``memory'' of its initial location in the flow.}}
  \end{figure}    
  
   \begin{figure*}
  \resizebox{\hsize}{!}{
   \includegraphics{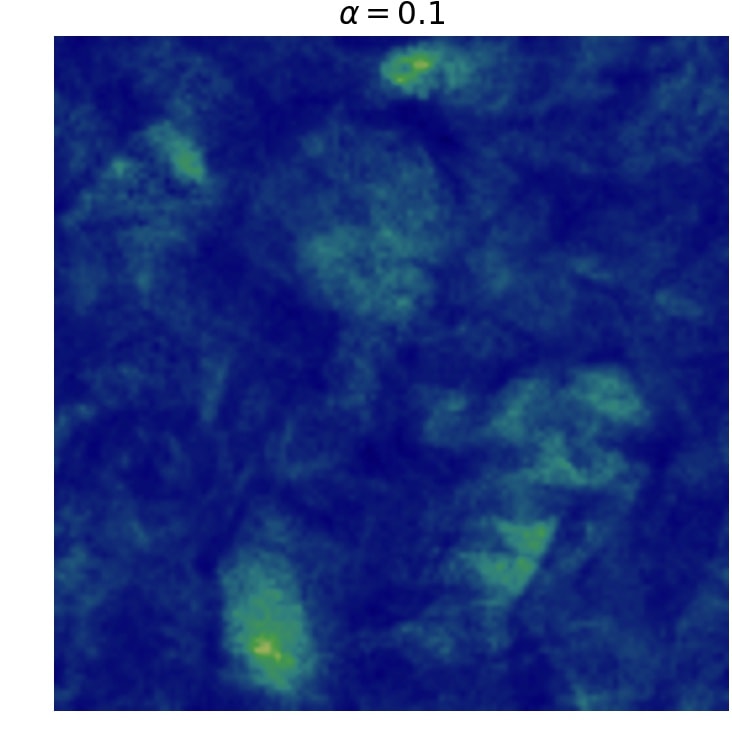}
   \includegraphics{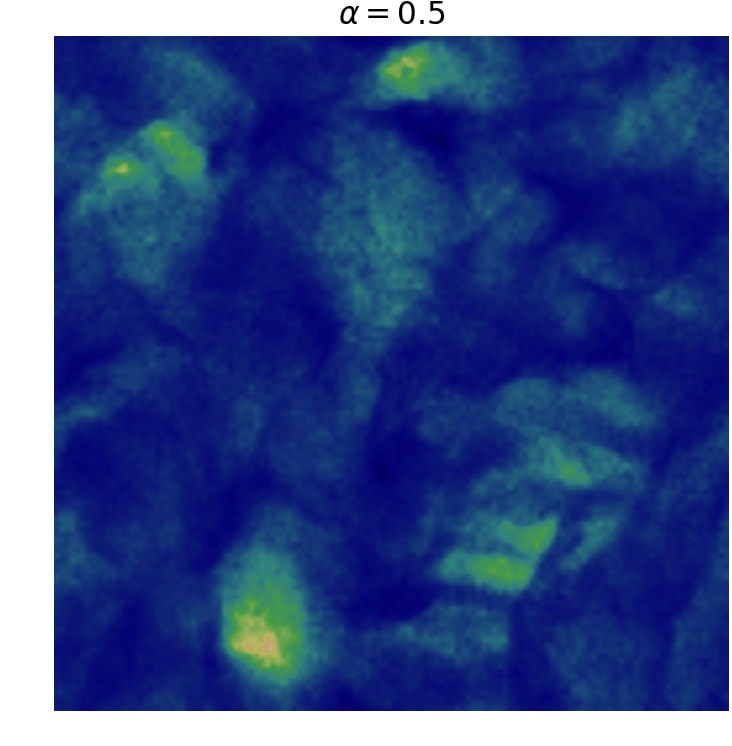}
   \includegraphics{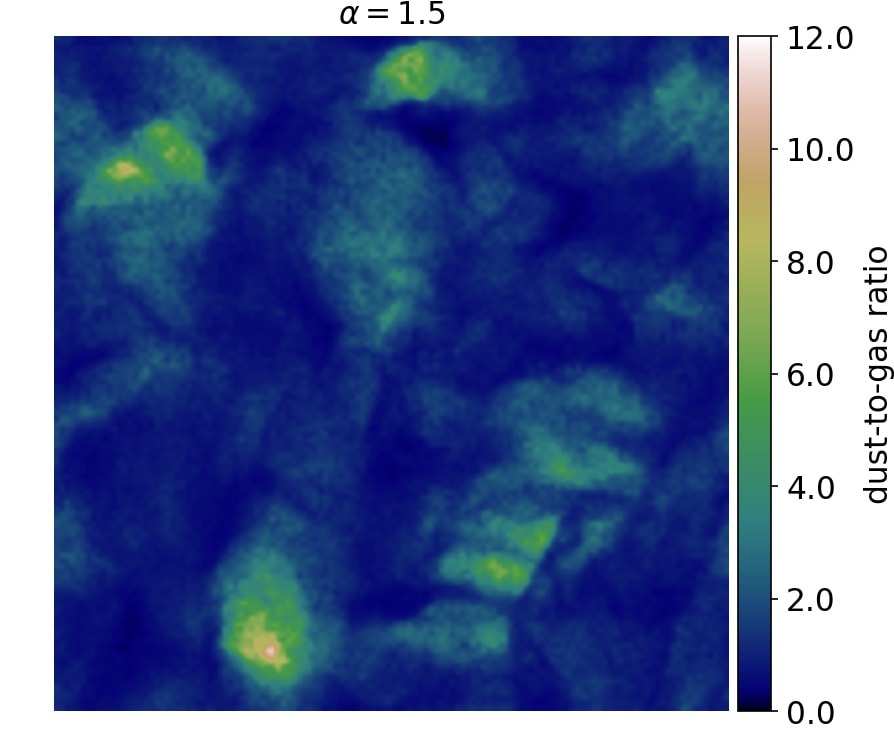}}
     \resizebox{\hsize}{!}{
   \includegraphics{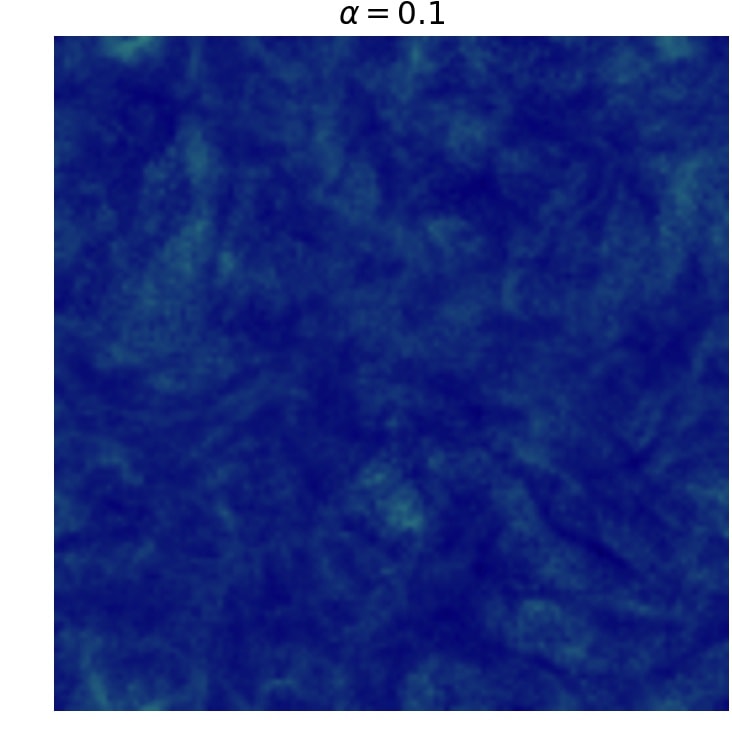}
   \includegraphics{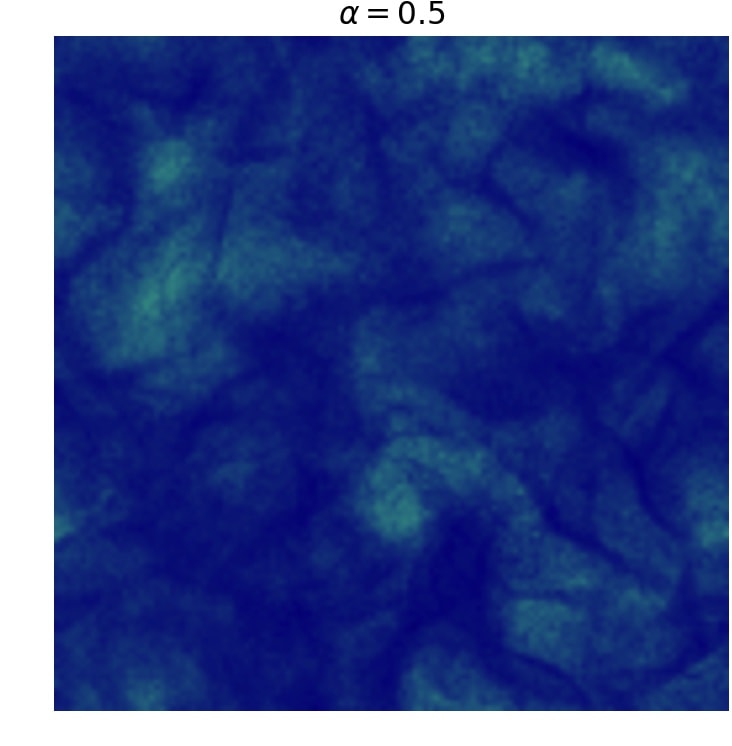}
   \includegraphics{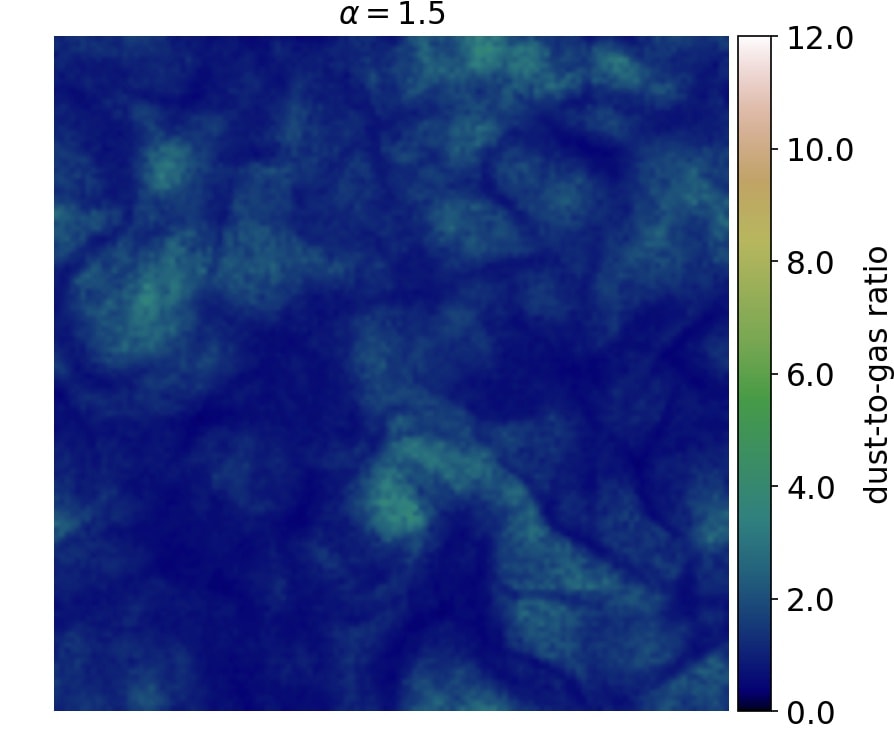}
   }
  \caption{\label{dtg} Projected dust-to-gas ratio for selected grain sizes ($\alpha$). Upper panels show results from simulation Bcmp, i.e., compressive forcing and a forcing factor $f=8.0$. Lower panels show results for the case of solenoidal forcing (Bsol). The average dust-to-gas ratio is normalised to 0.01 in all cases.}
  \end{figure*}

\section{Theory and background}
Incompressible (divergence-free) turbulence is characterised by how kinetic energy injected on large scales is transferred to successively smaller and smaller scales down to the viscous scales where it is dissipated. 
This kinetic-energy cascade is usually referred to as the Richardson cascade \citep{Richardson22}. In case of a compressible turbulent flow, however, the phenomenology becomes somewhat more complicated. The total kinetic energy does not provide a unified picture of compressible turbulence and there are several ways of accounting for compressibility that yield somewhat incompatible phenomenologies. 

\subsection{Governing equations}
\label{equations}
The basic equations governing the dynamics of the ISM are the equations of fluid dynamics (a.k.a. Navier-Stokes equations). For a compressible fluid/gas the density is given by the continuity equation,
\begin{equation}
{\partial \rho\over \partial t} + \nabla\cdot(\rho\,\mathbf{u})=0,
\end{equation}
where $\rho$ is density and $\mathbf{u}$ is the velocity field. The velocity field is governed by the momentum equation,
\begin{equation}
\label{EOM}
\rho\left({\partial \mathbf{u}\over \partial t} + \mathbf{u}\cdot\nabla\mathbf{u}\right)= -{\nabla P} +  {\mathbf{F}_{\rm visc}} + {\mathbf{F}_{\rm force}},
\end{equation}
in which $P$ is (gas) pressure, $\mathbf{F}_{\rm visc} = \nabla \cdot \,\left(2\nu\,\rho\,\mathbf {S} _{0}\right)+\nabla \cdot \,\left(3\zeta\, \mathbf {C} \right)$ represents viscous forces, where $\nu$ is the kinematic viscosity, $\mathbf{C}={1\over 3}\left(\nabla \!\cdot \!\mathbf {u} \right)\mathbf {I} $ is the compression tensor, $\mathbf{S}_0 = \mathbf{S} - \mathbf{C}$ is the rate-of-strain tensor, and $\mathbf{S} = {1\over 2}\left[\nabla \mathbf {v} +\left(\nabla \mathbf {v} \right)^{T}\right]$. The physical viscous forces are complemented with an artificial (shock) viscosity to ensure numerical stability. The last term in Eq. (\ref{EOM}), $\mathbf{F}_{\rm force}$, is external forcing, which in the present case represents stochastic driving of the turbulence. 

To obtain closure of the equations above we must also introduce a coupling between pressure and density, which is here just the isothermal condition $P=c_{\rm s}^2\,\rho$, with $c_{\rm s}$ the isothermal sound speed. 

\subsection{Forcing}
The last term in Eq. (\ref{EOM}), $\mathbf{F}_{\rm force}$, is stochastic forcing term with both solenoidal (rotational) and compressive components. The forcing is applied at low wave-numbers in Fourier space, following standard procedures. More precisely, the forcing is a white-in-time{\footnote{The stochastic variations represent white noise in the time domain.} stochastic process integrated using the Euler-Maruyama method, i.e., a stochastic differential equation (in Stratonovich form) is solved within the framework of It\^o calculus \citep{Revuz94}. In the present study we separate two physically different types of forcing; either purely compressive or purely solenoidal forcing (see Table \ref{simulations}) in order to explore if there will be any qualitative differences in the dust dynamics, since it is known that the resultant gas-density PDFs and the fractal properties of the gas are significantly different depending on whether the forcing is compressive or solenoidal \citep{Federrath09,Federrath10}. 

\subsection{Numerics, simulation setup and scaling}
\label{numerics}
We aim to model the cold ISM, e.g., the interior of a molecular cloud (MC) using a local, three-dimensional periodic-boundary box hydrodynamic model. That is, as described above, we solve the standard hydrodynamic equations (as described in Section \ref{equations}): the momentum equation, the continuity equation and an isothermal condition added as a closure relation, with a constant sound speed set to $c_{\rm s} = 1$. Dust particles are included as inertial particles in 10 size bins (see Fig. \ref{agrain}) with $10^6$ particles in each bin. In the present paper we will focus on three explicit sizes, however. These three correspond to relatively small grains (with a typical ISM scaling of the simulations corresponding to $a\sim 0.1\,\mu$m), intermediate-size particles ($a\sim 0.5\,\mu$m) and large particles ($a\sim 1.5\,\mu$m) that are expected to decouple from the flow. 

To solve the equations of the model, we use the Pencil Code, which is a non-conservative, high-order, finite-difference code (sixth order in space and third order in time) for compressible hydrodynamic flows with magnetic fields and particles. For a more detailed description of the code, see, e.g., \citet{Brandenburg02} and the Pencil Code website and GitHub page\footnote{http://pencil-code.nordita.org, https://github.com/pencil-code}. The simulations are performed in a three-dimensional periodic box with sides $L_x = L_y = L_z = 2\pi$ (dimensionless). Applied to the core of a MC, the physical size of the computational domain is roughly $L = 0.1-0.5\,$pc. That is, the resolution obtained in a $1024^3$ simulation is $\sim 20-100$~AU. For the initial state of the gas in the MC, we take a single thermal phase with constant number density, which in a real MC would correspond to a mean value between $10^3$ and $10^5$ cm$^{-3}$, or in mass density, roughly $10^{-21}$--$10^{-19}$ g\,cm$^{-3}$.

\subsection{Incompressible vs. compressible turbulence}
Much of the theory of particles in flows is based on studies of incompressible flows/fluids. For small Mach numbers the incompressible limit is a good approximation. However, the nature of astrophysical flows in general, and the dynamics of the ISM in particular, is such that we expect high Mach numbers and significant turbulence. Thus, we explore a new regime, for which current theory about particles in incompressible flows is inadequate to describe what happens with cosmic dust particles. 

\subsubsection{Incompressible turbulence}
The canonical description of incompressible turbulence in a purely hydrodynamic (Navier-Stokes) flow is due to \citet{Kolmogorov41}\footnote{The original article is in russian. However, an english translation is now available \citep{Kolmogorov91}.}, see also \citet{Fri96} for a modern introduction. 
The crucial idea is that the energy dissipation rate per-unit-volume, $\varepsilon$, is a constant even in the limit of zero viscosity. 
Dimensional arguments show that for velocity fluctuations across a length scale $\ell$, $v_{\ell} \sim \ell^{1/3}$, which implies that the shell-averaged energy in Fourier space goes as $E(k) \sim k^{-5/3}$ where $E(k)$ is the energy contained in shell in Fourier space of radius $k$. At present there is a large body of evidence from observations, experiments and direct numerical simulations that supports Kolmogorov's result~\footnote{Actually there are intermittency corrections to the Kolmogorov scaling laws but we ignore such small corrections in this context.}.

\subsubsection{Compressible turbulence}
Turbulent gases can be described in terms of the incompressible Navier-Stokes equations only if the root-mean-square value of the flow velocity is much smaller than the sound speed, i.e., at very low Mach numbers. This condition is rarely satisfied in astrophysical contexts and in particular not in the ISM. In the high Mach number limit, it is generally accepted (although not rigorously proven) that the three-dimensional Burgers' equation provides a good description of hypersonic turbulence \citep{Elmegreen04}. In such a case, if assuming a small but finite viscosity, it can be argued that the energy spectrum approaches $E(k)\propto k^{-2}$, which can be shown to derive from the energy spectrum being $E(k)\propto k^{-2}$ for a propagating step function \citep{Chuong10}. Since very high Mach numbers indicate that the flow is shock dominated, this energy spectrum mainly describes the compressible component and provides less information about the rotational structures (eddies) of a turbulent flow. Therefore, models of compressible turbulence are usually either a compromise between the low and high Mach-number limits or distinguish between energy spectra for the solenoidal and compressible components, respectively.

An attractive compromise is provided by the \citet{Fleck96} model. Building upon an idea proposed by \citet{vonWeizsacker51}, the Fleck model assumes there exist a self-similar hierarchy of sizes of cloud structures, which represents the density variations of a turbulent interstellar gas. This hierarchy is parameterised by a scaling exponent $\varepsilon$, $\rho \propto \ell^\varepsilon$, which is $\varepsilon = 0$ for the low Mach-number (incompressible) limit and $\varepsilon = 1$ for turbulent structures with perfect isotropic compression. \citet{Fleck96} also makes the assumption that the total rate of viscous dissipation is proportional to $\rho\,v\,\ell^{-1}$ in case of a compressible fluid. With this assumption, the self-similar hierarchy of structures predicts an energy spectrum of the form $E(k) \propto k^{-5/3 - 2\varepsilon}$. Obviously, with $\varepsilon = 0$ we obtain a spectrum of Kolmogorov type, while $\varepsilon = 1$ leads to a very steep energy spectrum $E(k) \propto k^{-11/3}$. Such a steep spectrum may not be realistic, although the spectrum can in principle be steeper than a Burgers spectrum ($\varepsilon = 1/6$). The simulations in the present study seem to suggest that compressible turbulence produce a spectrum which is at least as steep as the Burgers spectrum (see Fig. \ref{powerspectra}). We will return to this observation in Sect. \ref{kinspectra}.

\subsubsection{Clustering of particles in compressible turbulence}
Since compressible turbulence is expected to have a ``steeper-than-Kolmogorov'' energy spectrum, we have reasons to expect that clustering of particles (dust) embedded in the flow will be different. \citet{Nicolleau16} have demonstrated that changing the power index in a periodic kinematic simulation of turbulence affects the so-called clustering attractor of the particles. Established theories \citep[see, e.g.,][and references therein]{Monchaux12} for particles in incompressible turbulent flows can therefore not immediately be assumed to hold also for the highly compressible flows which are relevant for astrophysics. Thus, it is important to explore clustering of particles in compressible turbulence by direct numerical simulation (DNS).

\subsection{Dust grains in a turbulent flow}
\label{dustflow}
Assuming the dust is accelerated by the turbulent gas flow via an \citet{Epstein24} drag law, the equation of motion for dust particles embedded in the gas is 
\begin{equation}
\label{stokeseq}
{d \mathbf{v}\over d t}  = {\mathbf{u}-\mathbf{v}\over \tau_{\rm s}},
\end{equation}
where $\mathbf{v}$ and $\mathbf{u}$ are the velocities of the dust and the gas, respectively, and $\tau_{\rm s}$ is the so-called stopping time, i.e., the time it takes before a dust grain has accelerated (or decelerated) to same velocity as the gas flow (in case of a steady laminar flow). The stopping time in the Epstein limit depends on the size and density of the grain as well as the gas density and the relative Mach number $\mathcal{W}_{\rm s} = |\mathbf{u}-\mathbf{v}|/c_{\rm s}$ \citep{Schaaf63}. In the limit $\mathcal{W}_{\rm s}\ll 1$, we obtain 
\begin{equation}
\label{stoppingtime_incomp}
\tau_{\rm s}(\mathcal{W}_{\rm s}\ll 1) =  \sqrt{\pi\over 8}{\rho_{\rm gr}\over\rho}{a\over  c_{\rm s}} \equiv \tau_{\rm s,\,0},
\end{equation}
where $a$ is the grain radius (assuming spherical grains), $\rho_{\rm gr}$ is the bulk material density of the grain and the isothermal sound speed $c_{\rm s}$ replaces the thermal mean speed of molecules. The $\mathcal{W}_{\rm s}\ll 1$ case typically corresponds to small sonic Mach numbers, i.e., $\mathcal{M}_{\rm s}\ll 1$. For large $\mathcal{M}_{\rm s}$, we expect $\mathcal{W}_{\rm s}\gg 1$,
\begin{equation}
\label{stoppingtime_comp}
\tau_{\rm s} (\mathcal{W}_{\rm s}\gg 1)=  {4\over 3}{\rho_{\rm gr}\over\rho}{a\over  |\mathbf{u}-\mathbf{v}|}.
\end{equation}
Combining these two limits, we then obtain a convenient formula which is sufficiently accurate for our purposes \citep{Kwok75,Draine79}
\begin{equation}
\label{stoppingtime}
\tau_{\rm s} = \tau_{\rm s,\,0} \left(1 + {9\pi\over 128}{|\mathbf{u}-\mathbf{v}|^2\over c_{\rm s}^2 } \right)^{-1/2}
\end{equation}
The second term inside the parentesis can be seen as a correction for supersonic flow velocities and compression.

If the stopping time is much (several orders of magnitude) shorter than the characteristic timescale of the flow, which is the case for small (typically $a\lesssim 0.01\,\mu$m) dust particles in a dense gas ($n_{\rm H}\sim 10^4$~g~cm$^{-3}$), it is justified to make the simplification $\mathbf{v} = \mathbf{u}$. In the present paper, however, we are interested in the regime where this approximation does not hold and the dust grains decouple from the gas according to Eq. (\ref{stokeseq}).

We do not consider the ``back-reaction'' from dust grains, i.e., the drag effect an accelerated dust grain may have on the gas flow. The reason for this is that the dust-to-gas mass ratio and the mean speed of the dust particles are too small; there is simply not enough momentum in the dust phase to make a qualitative difference on a strongly forced turbulent gas flow. In case of radiative forcing on the grains, however, the situation is quite different and the drag that the dust exerts on the gas must be taken into account.

\subsection{Dimensionless quantities and other parameters}
\subsubsection{Flow variables}
In the isothermal case, the flow of the gas is characterised by two variables: $\rho$ and $\mathbf{u}$. Because the sound speed $c_{\rm s}$ is constant, it is natural to use $\mathbf{U}= \mathbf{u}/c_{\rm s}$ as a dimensionless simulation variable, and the density $\rho$ can be replaced with $s=\log(\rho/\langle \rho\rangle)$. The latter means that the density variable can be made dimensionless using an arbitrary reference density, which we chose to be the volumetric mean density $\langle \rho\rangle$. Similarly, we use $\mathbf{V}_i= \mathbf{v}_i/c_{\rm s}$ for the velocities of dust particles with size $i$, but there is no corresponding dust density variable since the dust grains are treated as discrete particles. As our simulations start from a uniform distribution of gas, initially at rest, it is convenient to choose $\rho(0)= \langle \rho\rangle$ as the unit density, i.e., $ \langle \rho\rangle =1$.  

With $\mathbf{U}$ and $s$ the dimensionless simulation variables describing the flow, we only need to set the time and length scales of our simulation. The length scale is most conveniently set to the size of the simulation box, i.e., length $L$ of one of the sides of the box. Because we use a periodic boundary condition, we chose $L=2\pi$. The time scale can then be chosen to be the sound-crossing time $\tau_{\rm sc}=L/c_{\rm s}$. With $c_{\rm s} = 1$, this means that $\tau_{\rm sc}=L = 2\pi$.

\subsubsection{Mach number}
Since we are simulating highly compressible flows, the Mach number, defined as the ratio of the flow speed and the sound speed, is an important number to characterise the flow; the higher the Mach number the higher the degree of compression. In the present paper we always refer to the root-mean-square of the sonic Mach number $\mathcal{M}_{\rm rms} = \sqrt{\langle \mathcal{M}_{\rm s}^2\rangle} = u_{\rm rms}\,c_{\rm s}^{-1}$ unless anything else is stated. Several other dimensionless quantities, such as the Reynolds number, can be expressed in terms of $\mathcal{M}_{\rm rms}$.

\subsubsection{Reynolds number and dimensionless viscosity}
Astrophysical flows are usually considered to be nearly inviscid and thus have very high Reynolds numbers (roughly Re~$>10^5$). In a simulation of forced turbulence the Reynolds number can be estimated from
\begin{equation}
{\rm Re} = {\bar{u}_{\rm rms}\over k_{\rm f}\,\nu} = c_{\rm s}\,{\mathcal{M}_{\rm rms} \over k_{\rm f}\,\nu},
\end{equation}
where $\bar{u}_{\rm rms}$ is the temporal average of the root-mean-square (rms) velocity (see time series of $u_{\rm rms}$; Fig. \ref{timeseries}) of the flow after a statistical steady state has been reached (all simulations rapidly reach a statistical steady state), $k_{\rm f}$ is the effective forcing wavenumber and $\nu$ is the kinematic viscosity. The rms Mach number of the simulations ranges between $\mathcal{M}_{\rm rms} = 3.24$ and $\mathcal{M}_{\rm rms} = 7.45$ (see Table \ref{simulations}) and the effective forcing wavenumber is $k_{\rm f}\approx 3$. The sound speed is unity, so the Reynolds numbers are given by  ${\rm Re}\approx q_{\rm s}\,\nu^{-1}$, where $q_{\rm s} = 1...2.5$ and $\nu$ is given in units of $c_{\rm s}$. To reach Re~$>10^5$ we would then need $\nu \lesssim 10^{-5}$, which is, unfortunately, not feasible for computational reasons. With the relatively strong forcing and moderate shock/artificial viscosity that we use, our simulations become stable only if $\nu \sim 5\cdot10^{-3}$ in case of a forcing factor $f=4$, $\nu \sim 0.01$ for $f=8$ and $\nu \sim 0.015$ for $f=12$, which means that the average Reynolds numbers are around Re$\,\sim 200$ in our simulations (see Table \ref{simulations}). This is clearly not a realistic value, but this issue is common for all simulations based on a finite difference scheme and a finite Re.

\subsubsection{Stokes number, Knudsen number and grain sizes}
The extent to which dust grains couple to the gas flow depends on two timescales: the time it takes for a grain to couple to the flow ($\tau_{\rm s}$, i.e., the stopping time described in Sect. \ref{dustflow}), and the time it takes before a given portion of the flow is reversed (a.k.a. the ``turnover time'' of the flow). The ratio between these timescales is a quantity which characterises the gas-grain interaction and is basically the definition of the Stokes number, ${\rm St} = {\tau_{\rm s}/ \tau_\ell}$. 

The Knudsen number ${\rm Kn} = \lambda_{\rm mfp}/a$, where $\lambda_{\rm mfp}$ is the mean-free path of the gas molecules, determines whether the kinetic drag can be described in the  fluid regime (${\rm Kn} \ll 1$) or the particle regime (${\rm Kn} \gg 1$). In the latter the stopping time has a simplified description sometimes referred to as ``Epstein drag'', which is particularly relevant for in an astrophysical context. Assuming also small relative Mach numbers, i.e., $\mathcal{W}\ll 1$, we then have
\begin{equation}
{\rm St}(\mathcal{W}\ll 1) \approx {\tau_{\rm s,\,0}\over \tau_\ell} = \sqrt{\pi\over 8}{\rho_{\rm gr}\over\rho}{u_{\rm rms}\over  c_{\rm s}}{a\over L} = \sqrt{\pi\over 8}{\rho_{\rm gr}\over\rho}\mathcal{M}_{\rm rms} {a\over L},
\end{equation}
where $\tau_{\ell}$ is the large-eddy turnover time $\tau_{\ell}\approx L\,u_{\rm rms}^{-1}$ and $L$ is the size of the simulation box. In the opposite limit ($\mathcal{W}\gg 1$) we have
\begin{equation}
{\rm St}(\mathcal{W}\gg 1) \approx  {4\over 3}{\rho_{\rm gr}\over\rho}{u_{\rm rms}\over  |\mathbf{u}- \mathbf{v}|}{a\over L} = {4\over 3}{\rho_{\rm gr}\over\rho}{\mathcal{M}_{\rm rms}\over \mathcal{W}} {a\over L}, 
\end{equation}
which means St is not a universal number. However, in the case of fully developed turbulence, $\tau_{\ell}$ is a statistical invariant and the Stokes number directly proportional to grain size. In principle, we could consider the volume averaged Stokes number $\langle {\rm St}\rangle$ as a measure of grain size in both limits of $\mathcal{W}$. But the scaling is different for different simulation setups, because $\langle {\rm St}\rangle$ depends on both $\mathcal{W}$ and the ordinary Mach number $\mathcal{M}_{\rm rms}$. Still, for $\mathcal{W}\ll 1$, it is possible to use $\langle {\rm St}\rangle/\mathcal{M}_{\rm rms}$ as a ``size parameter'', i.e., we may define
\begin{equation}
\alpha = {\rho_{\rm gr}\over\langle \rho\rangle}{a\over L} ,
\end{equation}
which is the parameterisation used  by \citet{Hopkins16}. However, because the total mass of a simulation box of size $L$ and the mass of a grain of a given radius $a$ are constants, the quantity $\alpha$ must also be a constant regardless of characteristics of the simulated flow. The parameter $\alpha$ is therefore a better dimensionless measure of grain size than the average Stokes number $\langle {\rm St}\rangle$ for a super-/hypersonic compressible flow. Following \citet{Hopkins16} the physical size of the grains can then be estimated from
\begin{equation}
a = 0.4\,\alpha\, \left({L\over 10\,{\rm pc}} \right)\left({\langle n_{\rm gas}\rangle \over 10\,{\rm cm}^{-3}} \right) \left({\rho_{\rm gr}\over 2.4\,{\rm g\,cm}^{-3}} \right)^{-1}\,\mu{\rm m},
\end{equation}
where $\langle n_{\rm gas}\rangle$ is the average number density of gas particles (molecules). Fig. \ref{agrain} shows a few examples of how the physical grain size scales with $\alpha$ depending on the adopted size of the simulation box and mean gas density.

\subsubsection{Dust-to-gas ratio}
As an indicator of dust-gas separation, we may consider another dimensionless quantity: the dust-to-gas ratio. If gas and dust is strongly coupled ($\mathbf{u} \approx \mathbf{v}$) this ratio shows very little spatial variation, while one would expect an anti-correlation with the gas if gas and dust are dynamically decoupled. Observationally, the dust-to-gas ratio is usually defined in terms of the dust mass density $\rho_{\rm d}$ instead of number density, i.e., the ratio obtained when the dust density is weighted by the grain-size distribution $\varphi(a)$ and the bulk material density of the dust,
\begin{equation}
\rho_{\rm d,\,\ell}(a) ={4\pi\over 3}{\rho_{\rm gr}}\,\langle n_{\rm d,\,\ell}\rangle\int_0^\infty a^3\,\varphi(a)\,da.
\end{equation}
For a mono-dispersed population of dust grains the expression simplifies into 
\begin{equation}
\rho_{\rm d,\,\ell}(a) ={4\pi\over 3}{\rho_{\rm gr}}\, a^3\,\langle n_{\rm d,\,\ell}\rangle.
\end{equation}
Moreover, the observed densities are column densities and observed dust-to-gas ratios are usually the ratio between the dust column and the gas column, i.e.,
\begin{equation}
Z_{\rm d}(a) = {\Sigma_{\rm d}(a)\over \Sigma } = \int_0^{L}\rho_{\rm d}(a,z)\,dz \,\, \Bigg / \, \int_0^{L}\rho(z)\,dz,
\end{equation}
where $L$ is the column depth (equal to the size of the simulation box in our case). This is the quantity plotted in Fig. \ref{dtg}, where we have also normailsed $Z_{\rm d}$ such that the average is $\langle Z_{\rm d}(a) \rangle = 0.01$ in all cases.

\subsubsection{Average nearest-neighbour ratio}
\label{NNS}
The dust-to-gas ratio does not say very much about the clustering of dust grains due to binning of the data necessary to obtain it. There are several way to quantify the clustering \citep{Monchaux12}, where one of the more direct approaches is nearest-neighbour statistics. Therefore we compute the first nearest-neighbour distance (1-NND) for each individual particle $i$ in each one of the simulations. We denote this parameter $\ell_{i,\,\rm m}$ and compare its ensemble average $\langle \ell_{i,\,\rm m} \rangle$  to the expected ensemble average in case of a random isotropic distribution of grains $\langle \ell_{i,\,\rm e}\rangle$, i.e., the case of no clustering. Then we calculate the average nearest neighbour (ANN) ratio, sometimes also called relative NND, as
\begin{equation}
R_{\rm ANN} = {\langle \ell_{i,\,\rm m}\rangle\over \langle\ell_{i,\,\rm e}\rangle} = 1.81\times {n_j^{1/3}\over N}{\sum_{i=1}^{i=N} \ell_{i,\,\rm m}},
\end{equation}
where $n_j$ is the number density of the considered particle species $j$ and the factor 1.81 comes from the normalision. The ANN ratio is a dimensionless measure of the clustering (or dispersion) of grains; if $R_{\rm ANN}<1$ the grains are clustered, while they can be regarded as dispersed if $R_{\rm ANN}>1$. The case $R_{\rm ANN}=1$ correspond to an exactly random distribution of grains. 

\subsubsection{Correlation dimension}
\label{D2}
Alternatively, we may consider a version of the 1-NND distribution which is parameterised in terms of the dimension-like parameter $D$,
\begin{equation}
h(D,r)\,dr=D\,\lambda(D)\,r^{D-1}\,g(D,r),
\end{equation}
and cumulative distribution
\begin{equation}
H(D,r)\,dr=\int_0^r h(D,r')\,dr',
\end{equation}
where $\lambda(D)$ is a parameter related to the volume of the $N$-sphere and $g(D,t)$ is a function with a bounded first derivative for all $\hat{r}\ge 0$ so that $r^D\,dg/d\hat{r}\to 0 $ as $r\to 0$ for all $D > 0$.  It can be demonstrated \citep[see][]{Mattsson18} that the parameter $D$ is equivalent to the so-called {\it correlation dimension} $d_2$ of the particle distribution (see Section \ref{D2_corrdim} for further details). In short, the method relies on the fact that the probability of finding a particle's {\it nearest neighbour} within a sphere of a radius $\hat{r}$ with the particle at its centre, is in fact proportional to the probability of finding {\it any} particle within that sphere. Consequently, one finds
\begin{equation}
d_2 \equiv {d\ln N_{\rm d}\over d\ln\hat{r}} = {d\ln H\over d\ln\hat{r}} = {h(D,\hat{r})\over \hat{r}\,H(D,\hat{r})} \approx D \quad {\rm for}\,\, \hat{r} \to 0.
\end{equation}
For the case $h(D,\hat{r})= \exp[{-\lambda(D)\,\hat{r}^D}]$, which has been suggested as an explicit general form for the 1-NND distribution \citep[see, e.g.,][]{Torquato90}, one obtains $d_2 = D$ exactly. $d_2<d$ corresponds to clustering in a $d$-dimensional spatial distribution of particles. Hence, fitting a power-law function to the left tail of the histogram of 1-NNDs is therefore a sufficiently reliable estimate\footnote{The correlation dimension can be computed with more direct methods, but such a procedure requires stacking of a large number of snapshots from a simulation and is computationally expensive.} of the correlation dimension $d_2$, which we will use in addition to the ANN to quantify clustering.

\section{Results and discussion}

\subsection{Power spectra}
\label{kinspectra}
\citet{Federrath13} has argued that the scaling with the wavenumber $k$ is steeper with compressive driving than with solenoidal driving. Hence, is the total "classical" energy spectrum of Kolmogorov or Burgers type, or something in between? The actual answer seems to be Burgers or even steeper (see Fig. \ref{powerspectra}), which is somewhat unexpected.

According to Kolmogorov's theory, we would expect that the rotational (parallel) component of the velocity field should yield essentially a Kolmogorov spectrum with $-5/3$ power-low slope. Similarly, there are reasons to believe that the compressive (transversal) component should be closer to a Burgers spectrum with a $-2$ slope. But the simulations with purely solenoidal forcing (Asol, Bsol and Csol in Table \ref{simulations}) seem to produce power spectra which are somewhat {\it steeper} than a Burgers spectrum and not slightly flatter spectra. 

For the simulations with compressive forcing (Acmp, Bcmp and Ccmp), the relatively high Mach numbers mean it is fair to anticipate a spectrum which is closer to the Burgers than the Kolmogorov case, at least within the well resolved part of the inertial range (approximately $4\le k\le 20$; cf. red line in Fig. \ref{powerspectra}). But it should be noted that strongly rotationally/solenoidally forced compressible flows are qualitatively different from the regime where the Kolmogorov theory is valid. 

\subsection{Density variations}
\subsubsection{Gas}
In the simulations with purely compressive forcing, the gas density PDFs for the logarithmic gas-density parameter $S=\log(\Sigma/\langle\Sigma\rangle)$ is very similar to a normal distribution at high densities, but show exponential tails for the low-density regime (see Fig. \ref{pdf_gas} for an example) in accordance with the results of \citet{Federrath08,Federrath10}. 

Turning to the simulations with purely solenoidal forcing (Asol, Bsol and Csol), we see two clear differences to the models with compressive forcing; there is no distinct low-density tail in the gas PDFs and the variances are smaller. The distribution of gas densities for the case of solenoidal forcing is also narrower than that for compressive forcing (see Fig. \ref{pdf_gas} for an example).

In more precise mathematical terms, we can describe the PDFs for $S$ in simulations with compressive forcing as a skewed lognormal distribution \citep{Azzalini85},
\begin{equation}
p(S) = {1\over\sqrt{2\pi}\,\omega}\left\{1+{\rm erf}\left[{\eta(S-\xi)\over \sqrt{2}\,\omega} \right] \right\}\,\exp\left[ -{(S-\xi)^2\over 2\,\omega^2}\right],
\end{equation}
where $\eta$, $\xi$ and $\omega$ are fitting parameters. With $\delta = \eta/\sqrt{1+\eta^2}$ we can write the mean, variance, skewness and kurtosis as
\begin{equation}
\langle S \rangle = \xi + \omega\delta\sqrt{2\over\pi},
\end{equation}
\begin{equation}
\sigma_S^2 = \omega^2 \left(1-{2\over\pi}\delta^2\right),
\end{equation}
\begin{equation}
\mathcal{S}_S = {4 -\pi\over 2}{(\delta\sqrt{2/\pi})^3\over (1-{2/\pi}\delta^2)^{3/2}},
\end{equation}
\begin{equation}
\mathcal{K}_S = {2(\pi-3)(\delta\sqrt{2/\pi})^4\over (1-{2/\pi}\delta^2)^{2}}.
\end{equation}
For simulations with solenoidal forcing the PDFs can be well fitted with an ordinary lognormal distribution, i.e.,
\begin{equation}
p(S) = {1\over\sqrt{2\pi}\,\sigma_S}\exp\left[ -{(S-\langle S \rangle)^2\over 2\,\sigma_S^2}\right],
\end{equation}
where mean, variance, skewness and kurtosis follows the usual moment hierarchy. In Table \ref{PDFfit} we list the mean, variance, skewness and kurtosis for the two examples (Bcmp and Bsol) in Fig. \ref{pdf_gas} and note that these quantities scale with the mean Mach number $\mathcal{M}_{\rm rms}$, as expected. However, for the simulations with compressive forcing the relation between variance and the Mach number is not the simple one expected when $S$ is normal distributed as in the case of solenoidal forcing. As the purpose of this paper is not to study the properties of the gas-density PDF, we will not consider the scalings with $\mathcal{M}_{\rm rms}$ in any detail. 

In Appendix \ref{gasdynamics} we show slices through the simulation box for logarithmic gas density and the local Mach number. Both these quantities show large variations and a structural correlation between them for both types of forcing. It is noteworthy that high Mach number and low gas density often correlate spatially in the simulations with compressive forcing, while this effect is not seen in the simulations with solenoidal forcing. In the latter, there is sooner a correlation  with high gas density. 

  \begin{table}
  \begin{center}
  \caption{\label{PDFfit} Resultant fitting parameters, mean and 1-$\sigma$ deviations from fitting of analytical the distribution functions to the gas-density PDFs obtained from simulations Bcmp (compressive forcing, $f = 8.0$) and Bsol (solenoidal forcing, $f = 8.0$).}
  \begin{tabular}{l|ll}
  \hline
  \hline
  \rule[-0.2cm]{0mm}{0.8cm}
   & Compressive  & Solenoidal \\
  \hline
   $\eta$ & -1.652 & -0.363 \\
   $\xi$   & 0.397  & 0.039 \\
   $\omega$ & 0.850 & 0.401 \\[2mm]
   $\langle S \rangle$ & -0.182 & 0.070 \\
   $\sigma_S$ & 0.621 & 0.386 \\
  \hline
  \hline
  \end{tabular}
  \end{center}
  \end{table}

\subsubsection{Dust}
As we decrease the size of the binning boxes, the dust-density PDFs seem to converge to a lognormal form for both types of forcing and all considered $\alpha$, as well as all sizes of the binning box $\ell$. In Fig. \ref{dustPDFs} we show the PDFs for $\alpha = $~0.1, 0.5, 1.5, with $\ell$ set to 1/16 of the resolution of the simulation box, as well as cumulative density functions (CDFs) for different binning-box sizes. The thick red (appears grey in printed version) curves in the CDF plots show estimates of the lognormal distribution these sequences seem to be converging towards (obtained by a rank-order technique). Estimating the PDF from CDFs is advantageous due to two unavoidable types of box-size biases. First, if the size of the binning box is not sufficiently small compered to the simulation box, small-scale structures are filtered out. Second, if the size of the binning box is close to the grid scale, we are seeing the effects of small-number statistics. (A reliable one-point PDF can only be obtained if a very large number of snapshots of a simulation is combined.) If the binning box is too large, we may also obtain an artificial power-law tail at the low-density end.

\subsection{Separation of gas and dust}
\label{separation}
Very small grains essentially behave like tracer particles and follow the gas flow, i.e., there is almost exact velocity coupling between the dust and gas phases. As grains grow bigger, and the Stokes number higher, the grains will tend to decouple more and more from the gas flow. In Figs. \ref{Ndust_Acmp}, \ref{Ndust_Bcmp} and \ref{Ndust_Ccmp}, showing the projected densities of gas and dust grains in the simulations with compressive forcing, one can clearly see that the smallest grains in our simulations ($\alpha= 0.1$) tend to end up more or less where the gas density is high, while larger grains show a spatial distribution which show little or no resemblance with the gas distribution and is generally random, isotropic and homogeneous. In our simulations with solenoidal forcing (Figs. \ref{Ndust_Asol}, \ref{Ndust_Bsol} and \ref{Ndust_Csol}) the correlation between dust and gas is generally stronger for the smallest grains, but also very clearly uncorrelated for large grains. It seems solenoidal forcing leads to more efficient decoupling and mixing of dust grains, which can be understood in terms of the angle between the velocity vector of a gas parcel and the velocity vector of a dust grain located in it, which is created due to rotational motion in combination with decoupling. Decoupling from vortices can lead to strong clustering of grains in between the vortices, which is probably why the smallest particles ($\alpha = 0.1$) appear more clustered than any of the larger particles in the simulations. If shock compression dominates, on the other hand, the velocity vectors may be of different magnitude, but the angle between them changes much less in each forcing ``kick'', which leads to less efficient mixing of the dust. These results are qualitatively identical to the findings of \citet{Hopkins16}, but in conflict with the results of \citet{Tricco17}\footnote{It should be noted that \citet{Tricco17} base their conclusion on smooth-particle hydrodynamics simulations, which require a different scheme for implementing kinetic drag on the dust. Whether this can explain the difference compared with \citet{Hopkins16} and the present study is unclear, however.}, which show essentially no significant separation for what appear to be similar $\alpha$-values.  

To quantify the (de)coupling between gas and dust, we introduce the coupling length: $\Delta x_{\rm c} = |\mathbf{u}-\mathbf{v}|\,\tau_{\rm s}$ (a.k.a. ``free-streaming length''). This quantity can vary significantly across the simulation box, so we will consider an approximation of the average,
\begin{equation}
\langle \Delta x_{\rm c}\rangle \approx 0.63\,\alpha\, c_{\rm s}\,\sqrt{\mathcal{W}_{\rm rms}^2\over 1 + 0.22\,\mathcal{W}_{\rm rms}^2},
\end{equation}
which measures how far a grain would typically be dislocated from its parent fluid (gas) element. That is, in Lagrangian coordinates, once a statistical steady state is reached, $\Delta x_{\rm c}$ is the typical distance between the parent fluid element of a grain at a given time and its location at any subsequent point in time. Here, we may note that in the limit of low relative Mach numbers ($\mathcal{W}_{\rm rms}\ll 1$), $\Delta x_{\rm c} \sim \alpha\,\mathcal{W}_{\rm rms}$, while in the opposite limit ($\mathcal{W}_{\rm rms}\gg 1$) $\Delta x_{\rm c} \sim \alpha$. The dependence on the mean relative Mach number $\mathcal{W}_{\rm rms}$ indicate a nonlinear relation since $\mathcal{W}_{\rm rms}$ is an increasing function of $a$, which is what we see for small $\alpha$ in Fig. \ref{clength}, where $\Delta x_{\rm c}$  relative to the size of the simulation box $L$ is plotted as a function of $\alpha$. From this figure it is evident that the simulations with solenoidal forcing have somewhat larger couplings lengths, which is also visible in a comparison of Figs. \ref{Ndust_Acmp}, \ref{Ndust_Bcmp} and \ref{Ndust_Ccmp} with Figs. \ref{Ndust_Asol}, \ref{Ndust_Bsol} and \ref{Ndust_Csol}. Moreover, it is noteworthy that in the size range $\alpha = 0.5\dots 1.0$, where dust grains start to show an almost random isotropic spatial distribution, the coupling length is $\Delta x_{\rm c}\sim L$, i.e., similar to the size of the simulation box.

The separation of gas and dust can also be seen in the dust-to-gas ratio $Z_{\rm d}$ for different $\alpha$. As shown in Fig. \ref{dtg} $Z_{\rm d}$ show less variation for the small grains with $\alpha = 0.1$. There is also clear difference between the simulations with compressive forcing relative to those with solenoidal; the latter show clearly less variation in $Z_{\rm d}$, indicating that the dust grains are better coupled to the gas (which is consistent with the shorter coupling lengths that we obtain). Nonetheless, all our simulations imply that dust and gas become increasingly uncorrelated with increasing grain size. This is expected, and it confirms the results by \citet{Hopkins16} presented in their Fig. 3. 

\subsection{Clustering of particles}
\label{D2_corrdim}
\citet{Bec07} adopted a method commonly used in molecular physics to measure the clustering of particles in terms of the correlation dimension, i.e., $d_2 = \lim_{\delta r\to 0} \left\{\ln[\langle \mathcal{N}(\delta r)\rangle]/\ln \delta r\right\}$, where $\mathcal{N}$ is the expected number of particles inside a ball of radius $\delta r$ surrounding a test particle \citep{Monchaux12,Gustavsson15}. \citet{Bec07} studied the concentrations of particles in simulated incompressible random/turbulent flows and found that $d_2$ reached a minimum at a Stokes number around ${\rm St}=0.7$, while approaching $d_2 = 3$ (no clustering) for large Stokes numbers (for very small Stokes numbers, the correlation dimension $d_2 \to 3$ as well, because the particles couple to the flow in that limit). 

We use the distribution of first nearest neighbour distance (1-NND) for all particles in our simulations to estimate $d_2$ as described in \citet{Mattsson18} and briefly outlined in Sect. \ref{NNS}. Furthermore, we also calculate the ANN ratio $R_{\rm ANN}$ to obtain a measure of how closely packed the particles are. The resultant numbers are given in Fig. \ref{corredim}. Clearly, there is significant clustering among the smaller particles ($\alpha = 0.1$), while the intermediate-size (moderately large) particles ($\alpha = 0.5$) are only weakly clustered and large particles ($\alpha = 1.5$) are essentially unclustered. The clustering of $\alpha =  0.1$ particles is likely due to the same small-scale clustering discussed by \citet{Bec07}, although the expected minimum of $d_2$ will likely occur at somewhat smaller $\alpha$ values than the ones considered here, which also means the minimum occurs at a lower $\alpha$ than in incompressible simulations \citep[e.g.,][]{Bhatnagar18}.

It is natural to attribute the shift of the $d_2$ minimum to the fact that we are simulating highly compressible flows, while \citet{Bec07} studied incompressible flows. In essence, the clustering of particles in compressible turbulence happens on two different scales. First, the compression of the gas means that dust particles coupled to the gas flow will be concentrated where the gas is. Second, on smaller scales, particles will cluster as a result of turbulent motions in the compressed gas.

As mentioned above, regarding small grains, the current simulations are not fully covering the expected dip in $d_2$ as a function of grain size. Scaled to the size and gas density of a typical molecular-cloud core the size range where this minimum likely occurs corresponds to nano dust particles ($a = 1-100$~nm). The smallest particles in our simulations ($\alpha = 0.1$) roughly correspond to grains of radius $a=100$~nm, assuming typical scaling parameters for the simulations. These grains are quite strongly clustered ($d_2=2.5 - 2.7$), but the minimum in $d_2$ is expected for $a<100$~nm \citep[which we believe is seen in the work of][]{Hopkins16}, suggesting a need for a follow-up study on the clustering of nano dust.

            \begin{figure*}
  \resizebox{\hsize}{!}{
   \includegraphics{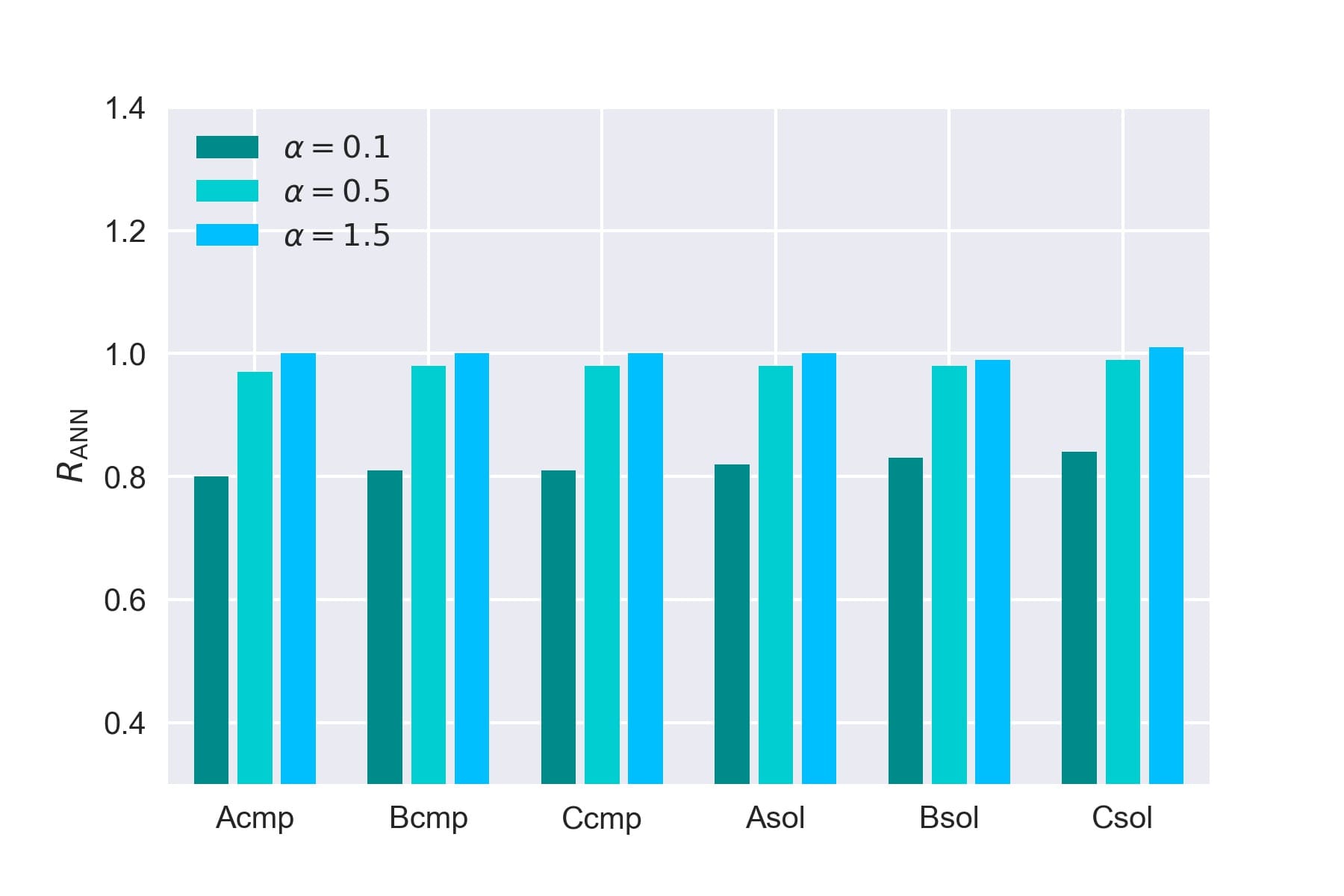}
   \includegraphics{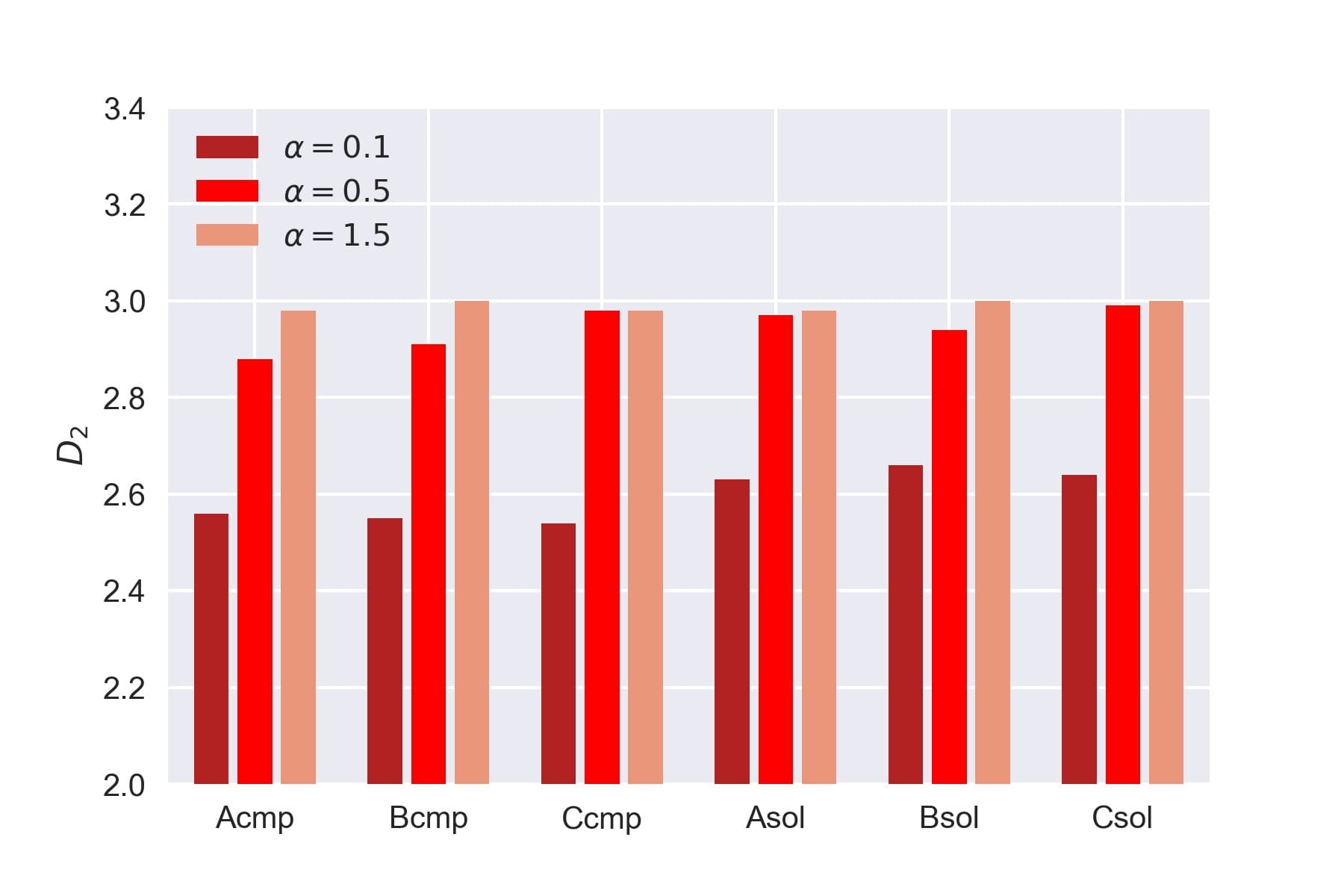}}
  \caption{\label{corredim} Quantitative analysis of grain clustering. Left: average first nearest-neighbour distance (ANN) ratios for three grain sizes ($\alpha$) for each simulation. Right: estimated correlation dimensions based on fits to the first nearest-neighbour distance (1-NND) distributions.}
  \end{figure*}

\subsection{Implications for gas heating}
\label{observables}
The result that dust and gas become increasingly uncorrelated with increasing grain size, and the fact that grains with different sizes tend to show a spatial displacement may have important effect on the heating of the ISM which is dominated by the photoelectric emission from dust grains \citep{Hollenbach99}.

As a matter of fact \citep[see, e.g.,][]{Weingartner01} the heating rate provided to the gas depends not only on the grain composition and charge state, but also on the grain size. More precisely, the photoelectric yields are enhanced for small grains \citep{Watson72} and therefore the heating efficiency decreases as the grain size increases. Consequently, aggregation/coagulation and fragmentation of grains (or any other process that reduces or increased the abundance of small grains) would affect the efficiency of gas heating. However, there is one more process to consider; the small-scale clustering of small dust grains seen in our simulations can, locally, lead to grain-size distributions biased towards small grains as well as subregions within a gas structure which are depleted in small grains. This means that estimates of the total photoelectric heating rate in a specific region are obviously sensitive to the grain size distribution in that specific region. Significant local variations in the clustering of dust grains of different size, as resulting from this work (see Figs. \ref{Ndust_Acmp} -- \ref{Ndust_Csol}), would therefore result in different heating in the various regions of the molecular cloud. Moreover, it should be noted that the spatial displacement of dust relative to gas might also affect the re-emission from interstellar dust grains, from the near-infrared to the microwave regime, as grains with different sizes have different optical properties.

\subsection{Implications for grain growth}
\subsubsection{Condensation (accretion of molecules)}
The growth velocity of a single grain with radius $a$ in a co-moving frame, $da/dt$, is proportional to the density of the relevant growth species $\rho_{i}$. More precisely, if the dust is tightly coupled to the gas flow, $da/dt = (8/\pi)^{1/2}f_{\rm s}\,c_{\rm s}\,\rho_{i}/\rho_{\rm gr}$, where $f_{\rm s}$ is the sticking probability and $\rho_{\rm gr}$ is the bulk density of the grain material. However, the growth velocity is affected by the separation of gas and dust. For large grains, which have a long average stopping time and thus experience more decoupling, condensation must be less efficient compared to small grains. The latter, on the other hand, represent much of the total grain-surface area and stay dynamically coupled to the gas to a much higher degree. The condensation rate is therefore not reduced by turbulence and the separation of gas and dust may not necessarily quench or even significantly lower the overall efficiency of dust condensation in turbulent MCs. But there exist a diffuse upper limit to sizes of grains grown by condensation. Once a grain has become large enough to completely decouple from the gas flow, i.e., when the stropping time $\tau_{\rm s}$ becomes comparable to the large-eddy timescale, $da/dt$ will decrease (on average) as the grain may not be located where the molecular gas density is high; a significant fraction of its lifetime the grain may reside in voids with very low gas density and therefore grow much slower.

\subsubsection{Coagulation (accretion of smaller grains)}
It is known that grain-growth by coagulation is enhanced by turbulent dynamics as the rate of grain-grain interaction $\Gamma_{ij}$ is proportional to velocity difference between the interacting particles of sizes $a_i$ and $a_j$. That is, $\Gamma_{ij}\propto \sigma_{ij}\,n_i\,n_j\,\Delta \mathbf{v}_{ij}$, where $\Delta \mathbf{v}_{ij}= |\mathbf{v}_i-\mathbf{v}_j|$ and $n_i$, $n_j$ are the number densities of particles of the considered sizes and $\sigma_{ij}$ is their total cross-section. Clustering of grains increases the interaction rate and thus the probability for coagulation irrespective of the dynamics of gas and dust since the number densities increase locally \citep{Mattsson16}. Grains of different sizes $\alpha$ will have different velocity distributions and show different degrees of clustering. But we have also seen that small grains and large grains do not seem to necessarily cluster at the same locations in the simulations, which casts doubt on the hypothesis that turbulent clustering is the main driving force behind turbulence-enhanced coagulation rates. The total cross-section $\sigma_{ij}$ is small for small particles and the number density of large grains is orders of magnitude lower than that for small grains, which means that if small and large grains are too efficiently separated due to turbulent gas motions and dynamical decoupling between gas and dust grains, it is highly unclear whether clustering plays the most important role. This is a complex problem that requires further study.

\subsection{Physics not included in the present simulations}
The simulations presented in the present paper are merely a first set of idealised simulations primarily intended to build a foundation for further research. The dynamics of particles in isothermal, purely hydrodynamic stochastically forced turbulence is theoretically quite well understood, especially in the incompressible limit. We limited the present paper to the study of how super/hypersonic turbulence arising from two fundamentally different types of forcing (compressive vs. solenoidal) affect dust particles of various sizes imbedded in the flow. Building upon these results, we may introduce more physics and study the effects of, e.g., relaxing the isothermal condition. A non-exhaustive list of physics needed to be considered in future work is presented here below.
\begin{itemize}
\item {\bf Self-gravity:} the gas-mass densities of MCs are high enough to have significant self-gravity effects if the ``physical'' size of the simulation box is comparable to (or larger than) the Jeans length. The combination of compressible turbulence and self-gravity is known to induce rapid clump formation and a high-density tail in the gas PDF \citep[see, e.g.,][]{Klessen00}. The dynamic decoupling of dust grains is affected by gravity and warrants further study \citep{Bhatnagar18}.
\item {\bf Magnetic fields:} the long-standing question of how turbulence amplify, sustain and shape magnetic fields can be turned around; what is the effect of magnetic forces on the turbulent gas dynamics of an interstellar MC? We have not considered the additional forces by magnetic fields that may be generated due to turbulence and the consequences it may have on the dynamics of dust grains. Statistically, however, there should be no major effect on neutral, non-magnetic grains. But future simulations should adress the problem of electrically charged grains with a magnetic dipole moments  and the qualitatively different dynamics of such grains due to Coulomb and Lorentz forces acting on the grains in addition to the kinetic drag force \citep{Draine03}. Due to charge fluctuations, i.e., the fact that even if neutral grains represent a significant part of the charge distribution of grains, those grains do not remain neutral a very long; charge fluctuations are usually so fast that one can assume that dust grains always carry a net average charge. Thus, magnetic fields will play a role under most circumstances and charge fluctuations should never be completely ignored \citep{Yan04}.
\item {\bf Shock heating of the gas:} in the present paper, all simulations assume an isothermal condition to obtain closure. Since interstellar gas is always highly compressible, i.e., the Mach numbers are high, any reasonable equation of state would yield an increase of temperature as the gas is compressed. Simulations involving heating/cooling and an entropy equation would be clearly more realistic as they would include local temperature variations. In this context, we should also mention that \citet{Kapyla18} found baroclinicity to be the most efficient vortex generator from supernovae (compressive forcing) such that with cooling processes included we would expect much more rotational flow even in highly compressible turbulence. 
\item {\bf Shock-destruction and accretion of molecules:} Hypersonic turbulence means strong shocks may form. Such shocks inside an MC can destroy dust in much the same way as the passage of a supernova shock. Depending on the type of forcing, magnetic fileds and gas density, the shocks may be dominated by ``continuous shocks'' (C-type) or ``jump shocks'' (J-type), which may both destroy dust, but inte slightly different ways \citep{Guillet07,Guillet09,Guillet11}. From an observational point of view, it is established that the abundance of certain molecules, e.g., SiO, could in some environments indicate shock destruction \citep{Savage96a}. Dust can of course also grow by accretion of molecules in MCs and including both destruction and growth/condensation of grains in simulations is important, since the balance between these two processes can be decisive for how the dynamics and clustering of grains develop.
\item {\bf Radiation:} we have assumed that thermal emission and absorption has negligible effects without the presence of stars. However, radiation pressure resulting from associations of hot stars forming inside MCs will have a profound effect on the dynamics of surrounding dust particles. 
\item {\bf Grain-grain interaction:} the collisional cross-section of the dust particles is effectively zero in our simulations. This means there is no scattering, coagulation or fragmentation due to grain-grain interaction taking place. But turbulence is expected to increase the interaction rate due to clustering and increased relative mean speed between interacting particles. These are highly localised phenomena and detailed simulations including simultaneous solution of the \citet{Smoluchowski16} equation are therefore an important step forward in our understanding of grain processing in MCs.
\item {\bf Back-reaction on the gas:} in the present work, we have only considered hydrodynamics drag on dust particles various sizes. If the dust mass contained in the gas is high enough, there will also be a back-reaction on the gas. Accelerated grains may exert a drag force on the gas, which will obviously affect velocity statistics as well as clustering. Simulations of interstellar dust particles in turbulent flows are usually based on the assumption that the back-reaction from the particles negligible. This assumption may not be valid in the presence of self-gravity or other mechanisms that may lead to locally high concentrations of dust grains. 
\end{itemize}

\section{Summary and conclusions}
We have performed 3D high resolution ($1024^3$) direct numerical simulations of stochastically forced (compressive as well as solenoidal) super/hypersonic turbulence with a multi-disperse population of (dust) particles imbedded in the flow. The resolution of our simulations is higher compared to many previous studies of this kind and the present study is probably the first that explicitly measures the correlation dimension of inertial particles in hypersonic turbulence. Moreover, we have also tested the effects of compressive vs. solenoidal forcing on grain dynamics and clustering. From our simulation results we conclude the following.
\begin{itemize}
\item The kinetic-energy power spectra are marginally consistent with a Burgers spectrum ($E(k)\propto k^{-2}$), where the simulations with purely compressive forcing are the ones that agree best with the Burgers case. Simulations with solenoidal forcing appear to be generally somewhat steeper. 
\item The gas PDFs of the simulations with solenoidal forcing is well approximated with a lognormal distributions, which is consistent with the results from several other authors. In case of purely compressive forcing, there seems to be low-density tails, again in agreement with previous work. The variance is also greater in case of compressive forcing and the PDF is best described with a relatively wide and skewed lognormal distribution. 
\item  The variation of the dust-to-gas ratio ($Z_{\rm d}$) depends on grain size and the type of forcing. Compressive forcing generates overall larger variance compared to solenoidal forcing. As grains decouple from the gas flow, variations of $Z_{\rm d}$ is mainly due to variations of the gas density.
\item Due to their longer stopping time, large grains ($\alpha \gtrsim 0.5$) will decouple from the gas flow, while small grains ($\alpha \lesssim 0.1$ or less) will tend to better trace the motions of the gas. This confirms the results by \citet{Hopkins16}, but raises also questions about the contrasting results by \citet{Tricco17}, which found that only very large grains show significant dynamic decoupling from the gas.
\item Simulations with purely solenoidal (as opposed to purely compressive) forcing show more dynamic decoupling of the larger dust grains ($\alpha \sim 1$), while the smallest grains in the simulations ($\alpha = 0.1$) appear to couple somewhat better to the gas flow than in the cases of compressive forcing. In general, the grains need to be smaller than the smallest grains in our simulations ($\alpha \lesssim 0.1$) in order to couple really well to the turbulent flows we simulate.
\item Numerical determination of the first nearest neighbour-distance (1-NND) distribution shows that smallest and the largest grains in our simulations display significant clustering. The correlation dimension $d_2$, used as a measure of clustering, is measured to be, for small grains ($\alpha \lesssim 0.1$), $d_2 \sim 2.5- 2.6$ in case of compressive forcing and $d_2 = 2.6 - 2.7$ in case of solenoidal forcing. For large grains ($\alpha \gtrsim 1$) $d_2$ is approaching the geometrical dimension ($d_2\approx 3$) for both compressive and solenoidal forcing. We have reason to believe that an expected minimum in $d_2$ as function of $\alpha$ will occur at smaller $\alpha$ values than we have considered in the present study. This warrants further study.
\item Small and large grains are not necessarily spatially correlated, which is why it is unclear whether the measured clustering will lead to significantly enhanced coagulation, although the grain-grain interaction rate of grains of similar sizes is locally elevated well above the average rate. It is also not obvious how growth by condensation is affected by turbulence in combination with dynamic decoupling of grains, but we expect there to be a diffuse upper limit to the grain sizes reached by condensational growth, which is due to the fact that large grains may not be located where the molecular gas is. 
\end{itemize}

\section*{Acknowledgments}
The anonymous reviewer is thanked for his/her exemplarily concise and constructive comments, which helped improving the paper. Andrea Bracco is thanked for his valuable comments on a draft version of this paper. Bengt Gustafsson is thanked for numerous thought-provoking discussions related to this work and for his insightful comments in general. Finally, we would like to give very special acknowledgements to Dhrubaditya Mitra and Livia Vallini for their highly valuable contributions and support. This project is supported by the Swedish Research Council (Vetenskapsrådet, grant no. 2015-04505) and by the grant ``Bottlenecks for particle growth in turbulent aerosols'' from the Knut and Alice Wallenberg Foundation (Dnr. KAW 2014.0048). Nordita is financed by the Nordic Council of Ministers and the two host universities KTH Royal Institute of Technology and Stockholm University.

\bibliographystyle{mnras}
\bibliography{refs_dust}

\appendix
\onecolumn
\section{Properties of the gas flow}
\label{gasdynamics}
In Fig. \ref{Machandcheese} we present snapshots of $x$--$y$ slices (one grid cell in thickness) through the middle of the simulation boxes for the logarithmic gas-density parameter $s$ (left panels) and the local sonic Mach number $\mathcal{M}_{\rm s}$ (right panels). High Mach numbers sometimes correlate with voids and in other cases with density peaks. The same is true for low Mach numbers. This is expected in hypersonic turbulence, because density peaks build up only after a shock has formed. Whether the density is high or low where the Mach number is high depends on in which stage a shock has been captured  Furthermore, we see arc-like density structures forming as a consequence of propagating and interacting shocks (see Fig. \ref{Machandcheese}, left panels). These gas-density structures have in many cases corresponding structures in the convergence of the velocity field. Although the dynamics of the gas is clearly shock dominated, there is a significant rotational part of the flow as well (see Fig. \ref{columns_divrot}, right panels, showing the projected absolute value of the vorticity). It is interesting to note that the strength of shocks and rotation of the flow tend to correlate spatially (compare the left and right panels in Fig. \ref{columns_divrot}). 

        \begin{figure*}
  \resizebox{0.90\hsize}{!}{ 
    \includegraphics{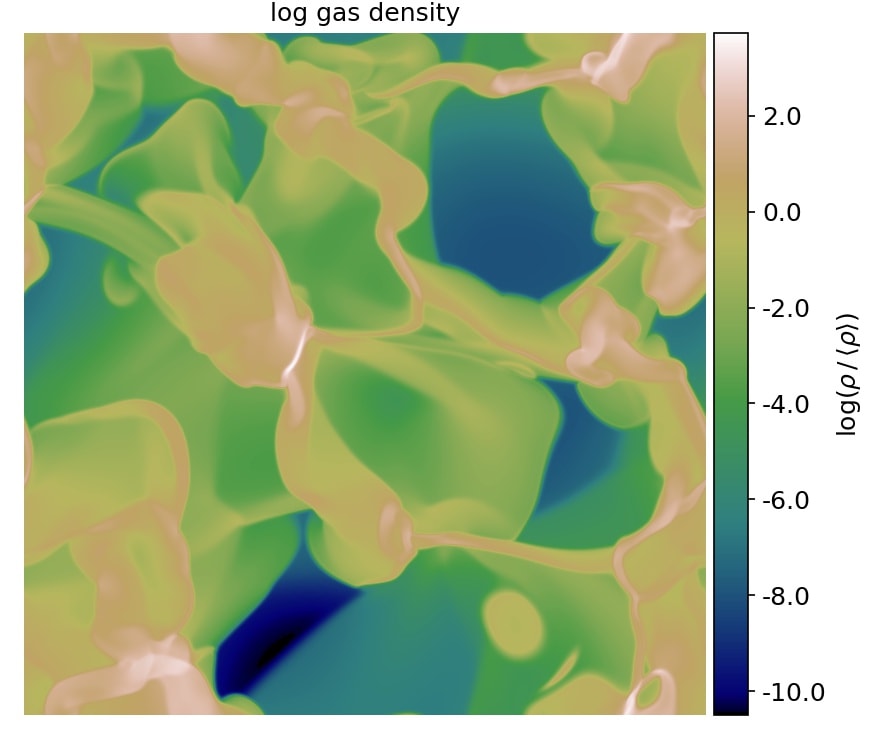}
    \includegraphics{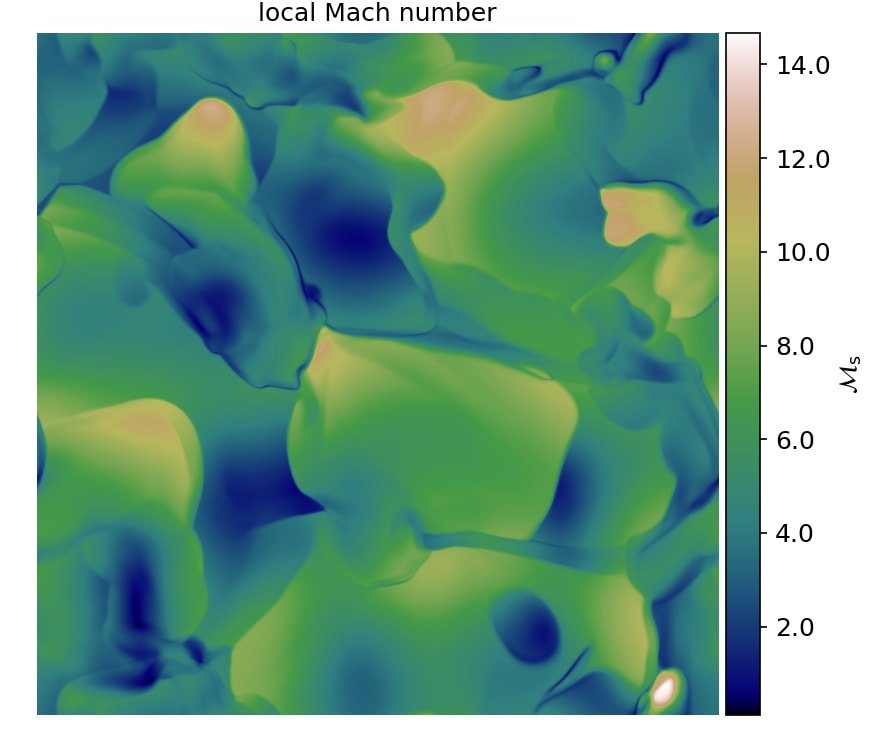}}
  \resizebox{0.90\hsize}{!}{ 
    \includegraphics{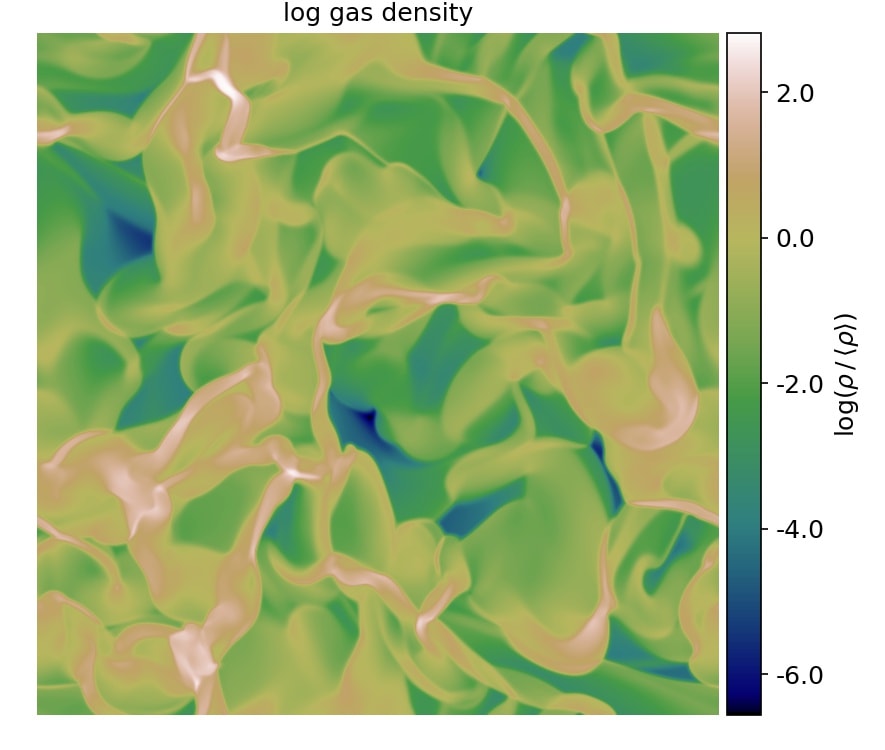}
    \includegraphics{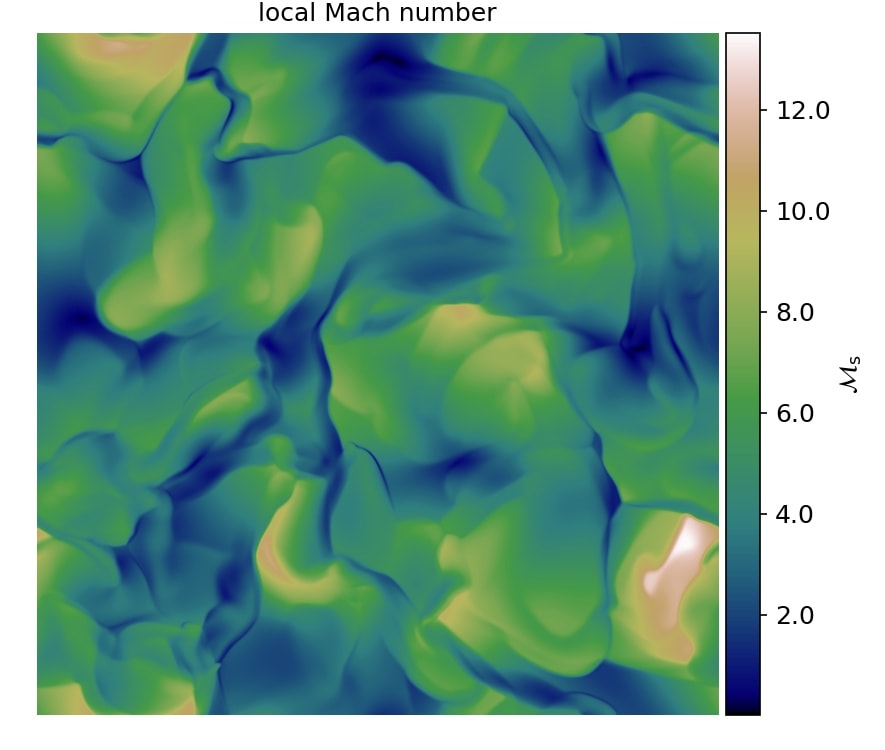}}
  \caption{\label{Machandcheese} Left panels: logarithm of gas density in slice of the simulation box taken through the middle of the box for models with purely compressive forcing (Bcmp; upper panel) and purely solenoidal (Bsol; lower panel). Right panels: the local  Mach number ($\mathcal{M}_{\rm s} = |\mathbf{u}|/c_{\rm s}$) in the same slice. Both simulations assume a forcing parameter of $f=8.0$ and the slices are obtained from snapshots taken at the end of the time series.}
  \end{figure*}

As expected, purely solenoidal forcing generates slightly different results compared to compressive forcing. The slice plots of gas density and Mach numbers show more filamentary structure and the range of gas densities are generally smaller (Fig. \ref{pdf_gas}) while slightly higher maximum Mach numbers are seen with solenoidal forcing, compared to compressive forcing (see Table \ref{simulations} and Fig. \ref{Machandcheese}, blue line); the total kinetic energy of the flow must be roughly the same in both cases. Thus, the larger spread in gas densities obtained with purely compressive forcing must be compensated with a smaller range of Mach numbers compared to a case with purely solenoidal forcing. The shock compression is generally the same in simulations with the same forcing parameter regardless of the type forcing (compressive or solenoidal), but the vorticity is stronger in the simulations with solenoidal forcing (see Fig. \ref{columns_divrot} for an example), since ``stirring'' will obviously generate more rotation. There is also some difference in how the vorticity is distributed spatially, which is interesting since in our current understanding of clustering of particles in turbulence vorticity is essential to explain the phenomenon\footnote{Rotation in the gas is centrifuging of particles away from vortex cores leading to accumulation of particles in convergence zones in between vortices.} \citep[see][and references therein]{Toschi09}.

        \begin{figure*}
     \resizebox{0.90\hsize}{!}{
   \includegraphics{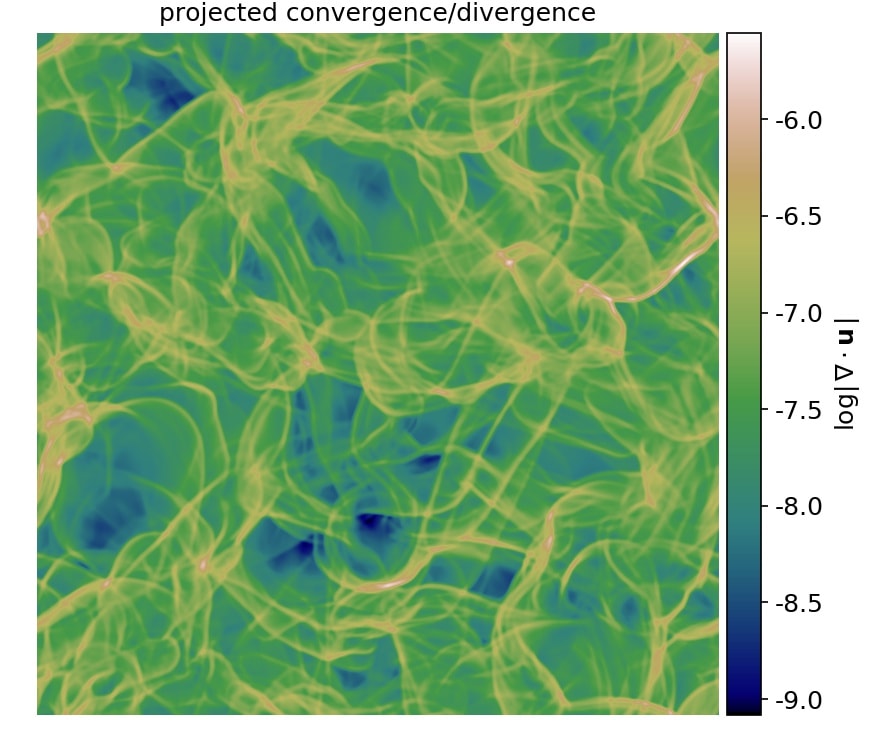}
   \includegraphics{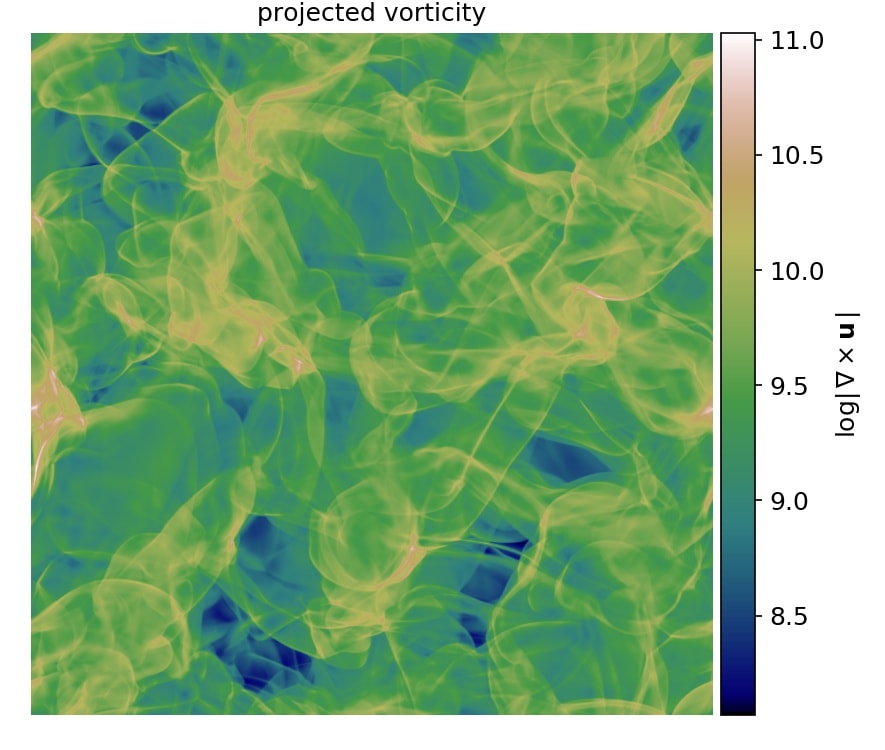}}
     \resizebox{0.90\hsize}{!}{
   \includegraphics{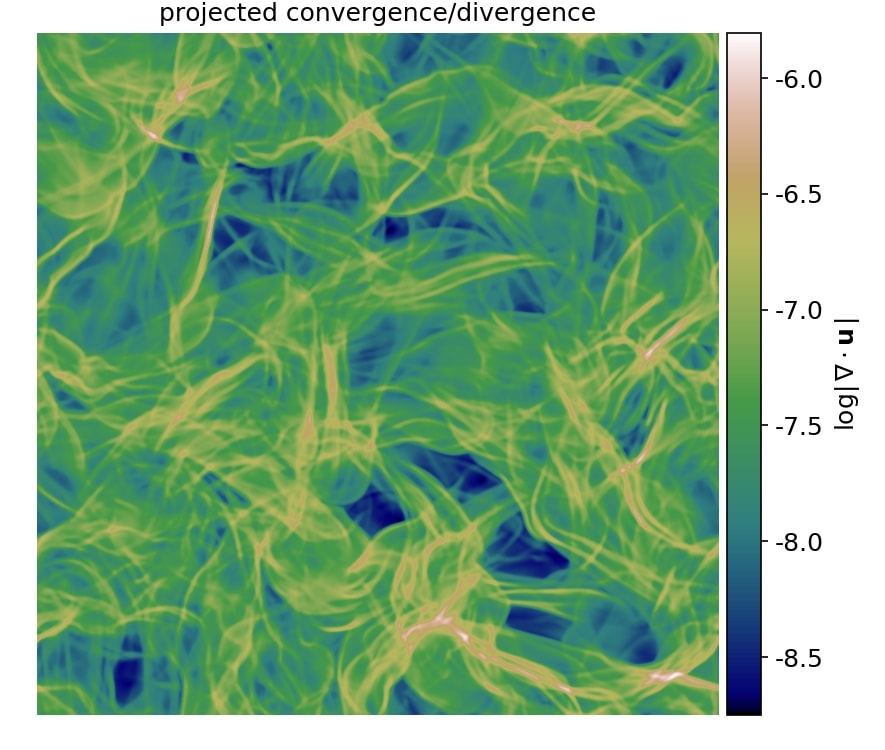}
   \includegraphics{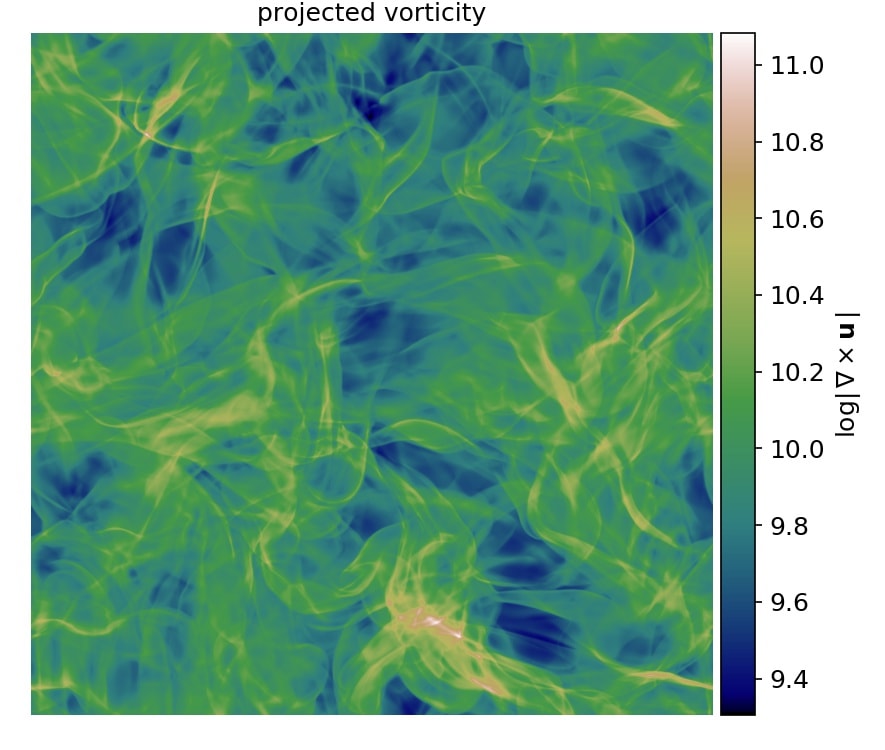}}

  \caption{\label{columns_divrot} Left panels: projected ``shock density'' (absolute value of the divergence of the velocity field) for simulations Bcmp (upper) and Bsol (lower) with purely compressive and solenoidal forcing, respectively, and a forcing parameter $f=8.0$. Right panels: projected absolute value of vorticity (rotation) for the same two simulations. The projections are calculated from snapshots taken at the end of the time series.}
  \end{figure*}

\end{document}